\documentclass[twocol]{ametsocV6.1}

\usepackage{amsmath,amsfonts,amssymb,bm}
\usepackage{mathptmx}
\usepackage{newtxtext}
\usepackage{newtxmath}
\usepackage{nicefrac}

\newcommand{\abs}[1]{\left \lvert #1 \right\rvert} 

\renewcommand{\vec}[1]{\boldsymbol{#1}  } 

\renewcommand{\d}[2]{\frac{\mathrm{d} #1}{\mathrm{d} #2}} 
\newcommand{\dd}[2]{\frac{\mathrm{d}^2 #1}{\mathrm{d} #2^2}} 
\newcommand{\pd}[2]{\frac{\partial #1}{\partial #2}} 
\newcommand{\grad}[1]{\vec{\nabla} #1} 
\newcommand{\lap}{\nabla^2} 
\newcommand{\unit}[1]{\hat{\vec{#1}}}

\newcommand{\e}{\mathrm{e}}
\newcommand{\irm}{\mathrm{i}}
\newcommand{\di}[1]{\mathrm{d}#1}

\newcommand{\Esc}{\mathscr{E}}

\newcommand{\Ssc}{\mathscr{S}}
\newcommand{\Ksc}{\mathscr{K}}
\newcommand{\Psc}{\mathscr{P}}

\newcommand{\J}[2]{ \mathrm{J}\left(#1,#2\right) }

\newcommand{\psik}{\hat \psi_{\vec k}}

\newcommand{\thetak}{\hat \theta_{\vec k}}

\title{Surface Quasigeostrophic Turbulence in Variable Stratification}

\authors{Houssam Yassin,\aff{a}\correspondingauthor{Houssam Yassin, hyassin@princeton.edu} 
Stephen M. Griffies,\aff{a,b}
}

\affiliation{\aff{a}{Program in Atmospheric and Oceanic Sciences, Princeton University, Princeton, NJ, USA}\\
\aff{b}{NOAA/Geophysical Fluid Dynamics Laboratory, Princeton, NJ, USA}\\
}

\abstract{Numerical and observational evidence indicates that, in regions where mixed-layer instability is active, the surface geostrophic velocity is largely induced by surface buoyancy anomalies. Yet, in these regions, the observed surface kinetic energy spectrum is steeper than predicted by uniformly stratified surface quasigeostrophic theory. By generalizing surface quasigeostrophic theory to account for variable stratification, we show that surface buoyancy anomalies can generate a variety of dynamical regimes depending on the stratification's vertical structure. Buoyancy anomalies generate longer range velocity fields over decreasing stratification and shorter range velocity fields  over increasing stratification. As a result, the surface kinetic energy spectrum is steeper over decreasing stratification than over increasing stratification. An exception occurs if the near surface stratification is much larger than the deep ocean stratification. In this case, we find an extremely local turbulent regime with surface buoyancy homogenization and a steep surface kinetic energy spectrum, similar to equivalent barotropic turbulence. By applying the variable stratification theory to the wintertime North Atlantic, and assuming that mixed-layer instability acts as a narrowband small-scale surface buoyancy forcing, we obtain a predicted surface kinetic energy spectrum between $k^{-4/3}$ and $k^{-7/3}$, which is consistent with the observed wintertime $k^{-2}$ spectrum. We conclude by suggesting a method of measuring the buoyancy frequency's vertical structure using satellite observations.} 

\begin{document}

\maketitle

%
%
%
%
%


\section{Introduction}\label{S-intro}

\subsection{Geostrophic flow induced by surface buoyancy}

Geostrophic flow in the upper ocean is induced by either interior potential vorticity anomalies, $q$, or surface buoyancy anomalies, $b|_{z=0}$.
At first, it was assumed that the surface geostrophic flow observed by satellite altimeters is due to interior potential vorticity \citep{stammer_global_1997,wunsch_vertical_1997}.
It was later realized, however, that under certain conditions, upper ocean geostrophic flow can be inferred using the surface buoyancy anomaly alone \citep{lapeyre_dynamics_2006,lacasce_estimating_2006}. 
Subsequently, \cite{lapeyre_what_2009} used a numerical ocean model to show that the surface buoyancy induced geostrophic flow dominates the $q$-induced geostrophic flow over a large fraction of the North Atlantic in January.
Lapeyre then concluded that the geostrophic velocity inferred from satellite altimeters in the North Atlantic must usually be due to surface buoyancy anomalies instead of interior potential vorticity.

Similar conclusions have been reached in later numerical studies using the effective surface quasigeostrophic  \citep[eSQG, ][]{lapeyre_dynamics_2006} method. The eSQG method aims to reconstruct three-dimensional velocity fields in the upper ocean: it assumes that surface buoyancy anomalies generate an exponentially decaying streamfunction with a vertical attenuation determined by the buoyancy frequency, as in the uniformly stratified surface quasigeostrophic model \citep{held_surface_1995}.
 Because the upper ocean does not typically have uniform stratification, an "effective" buoyancy frequency is used, which is also intended to account for interior potential vorticity anomalies \citep[][]{lapeyre_dynamics_2006}. 
In practice, however, this effective buoyancy frequency is chosen to be the vertical average of the buoyancy frequency in the upper ocean. 
 \cite{qiu_reconstructability_2016} derived the surface streamfunction from sea surface height in a $\nicefrac{1}{30}^\circ$ model of the Kuroshio Extension region in the North Pacific and used the eSQG method to reconstruct the three-dimensional vorticity field. They found correlations of 0.7-0.9 in the upper 1000 m between the reconstructed and model vorticity throughout the year. This result was also found to hold in a $\nicefrac{1}{48}^\circ$ model with tidal forcing \citep{qiu_reconstructing_2020}.


A clearer test of whether the surface flow is induced by surface buoyancy is to reconstruct the geostrophic flow directly using the sea surface buoyancy or temperature \citep{isernfontanet_potential_2006}. 
This approach was taken by \cite{isernfontanet_three-dimensional_2008} in the context of a $\nicefrac{1}{10}^\circ$ numerical simulation of the North Atlantic. 
When the geostrophic velocity is reconstructed using sea surface temperature, correlations between the reconstructed velocity and the model velocity exceeded 0.7 over most of the North Atlantic in January. 
Subsequently, \cite{miracca-lage_can_2022} used a reanalysis product with a grid spacing of 10 km to reconstruct the geostrophic velocity using both sea surface buoyancy and temperature over certain regions in the South Atlantic. 
The correlations between the reconstructed streamfunctions and the model streamfunction had a seasonal dependence, with correlations of 0.7-0.8 in winter and 0.2-0.4 in summer. 

Observations also support the conclusion that a significant portion of the surface geostrophic flow may be due to surface buoyancy anomalies over a substantial fraction of the World Ocean.
 \cite{gonzalez-haro_global_2014} reconstructed the surface streamfunction using $\nicefrac{1}{3}^\circ$ satellite altimeter data (for sea surface height) and $\nicefrac{1}{4}^\circ$ microwave radiometer data (for sea surface temperature). 
If the surface geostrophic velocity is due to sea surface temperature alone, then the streamfunction constructed from sea surface temperature should be perfectly correlated with the streamfunction constructed from sea surface height. 
The spatial correlations between the two streamfunctions was found to be seasonal. 
For the wintertime northern hemisphere, high correlations (exceeding 0.7-0.8) are observed near the Gulf Stream and Kuroshio whereas lower correlations (0.5-0.6) are seen in the eastern recirculating branch of North Atlantic and North Pacific gyres [a similar pattern was found by \cite{isernfontanet_three-dimensional_2008} and \cite{lapeyre_what_2009}]. In summer, correlations over the North Atlantic and North Pacific plummet to 0.2-0.5, again with lower correlations in the recirculating branch of the gyres.
In contrast to the strong seasonality observed in the northern hemisphere, correlation over the Southern Ocean typically remain larger than 0.8 throughout the year.

Another finding is that more of the surface geostrophic flow is due to surface buoyancy anomalies in regions with high eddy kinetic energy, strong thermal gradients, and deep mixed layers \citep{isernfontanet_three-dimensional_2008, gonzalez-haro_global_2014, miracca-lage_can_2022}. 
These are the same conditions under which we expect mixed-layer baroclinic instability to be active \citep{boccaletti_mixed_2007,mensa_seasonality_2013,sasaki_impact_2014,callies_seasonality_2015}. Indeed, one model of mixed-layer instability consists of surface buoyancy anomalies interacting with interior potential vorticity anomalies at the base of the mixed-layer \citep{Callies2016}. 
The dominance of the surface buoyancy induced velocity in regions of mixed-layer instability suggests that, to a first approximation, we can think of mixed-layer instability as energizing the surface buoyancy induced part of the flow through vertical buoyancy fluxes and the concomitant release of kinetic energy at smaller scales. 


\subsection{Surface quasigeostrophy in uniform stratification}

The dominance of the surface buoyancy induced velocity suggests that a useful model for upper ocean geostrophic dynamics is the surface quasigeostrophic model \citep{held_surface_1995,lapeyre_surface_2017}, which describes the dynamics induced by surface buoyancy anomalies over uniform stratification.
 The primary difference between surface quasigeostrophic dynamics and two-dimensional barotropic dynamics \citep{kraichnan_inertial_1967} is that surface quasigeostrophic eddies have a shorter interaction range than their two-dimensional barotropic counterparts. 
 One consequence of this shorter interaction range is a flatter kinetic energy spectrum \citep{pierrehumbert_spectra_1994}. Letting $k$ be the horizontal wavenumber, then two-dimensional barotropic turbulence theory predicts a kinetic energy spectrum of $k^{-5/3}$ upscale of small-scale forcing and a $k^{-3}$ spectrum downscale of large-scale forcing \citep{kraichnan_inertial_1967}. 
 If both types of forcing are present, then we expect a spectrum between $k^{-5/3}$ and $k^{-3}$, with the realized spectrum depending on the relative magnitude of small-scale to large-scale forcing \citep{lilly_two-dimensional_1989,maltrud_energy_1991}. 
 In contrast, the corresponding spectra for surface quasigeostrophic turbulence are $k^{-1}$ (upscale of small-scale forcing) and $k^{-5/3}$ (downscale of large-scale forcing) \citep{blumen_uniform_1978}, both of which are flatter than the corresponding two-dimensional barotropic spectra.\footnote{The uniformly stratified geostrophic turbulence theory of \cite{charney_geostrophic_1971} provides spectral predictions similar to the two-dimensional barotropic theory \citep[See ][]{callies_interpreting_2013}.} 
 
The above considerations lead to the first discrepancy between the surface quasigeostrophic model and ocean observations. 
As we have seen, we expect wintertime surface geostrophic velocities near major extratropical currents to be primarily due to surface buoyancy anomalies.
Therefore, the predictions of surface quasigeostrophic theory should hold.
If we assume that mesoscale baroclinic instability acts as a large-scale forcing and that mixed-layer baroclinic instability acts as a small-scale forcing to the upper ocean \citep[we assume a narrowband forcing in both cases, although this may not be the case, see][]{khatri_role_2021}, then we expect a surface kinetic energy spectrum between $k^{-1}$ and $k^{-5/3}$.
However, both observations and numerical simulations of the Gulf Stream and Kuroshio find kinetic energy spectra close to $k^{-2}$ in winter  \citep{sasaki_impact_2014,callies_seasonality_2015,vergara_revised_2019}, which is steeper than predicted.


The second discrepancy relates to the surface transfer function implied by uniformly stratified surface quasigeostrophic theory. The surface transfer function, $\mathcal{F}(\vec k)$, is defined as \citep{isern-fontanet_transfer_2014}
\begin{equation}\label{eq:transfer_func}
	\mathcal{F}(\vec k) = \frac{\hat \psi_{\vec k}}{\hat b_{\vec k}},
\end{equation}
where $\hat \psi_{\vec k}$ and $\hat b_{\vec k}$ are the Fourier amplitudes of the geostrophic streamfunction, $\psi$, and the buoyancy, $b$, at the ocean's surface, and $\vec k$ is the horizontal wavevector. Uniformly stratified surface quasigeostrophic theory predicts an isotropic transfer function $\mathcal{F}(k)\sim k^{-1}$ \citep{held_surface_1995}. Using a $\nicefrac{1}{12}^\circ$ ocean model and focusing on the western coast of Australia, \cite{gonzalez-haro_ocean_2020} confirmed that the transfer function between sea surface temperature and sea surface height is indeed isotropic but found that the transfer function is generally steeper than $k^{-1}$. In another study using a $\nicefrac{1}{16}^\circ$ model of the Mediterranean Sea, \cite{isern-fontanet_transfer_2014} found that the transfer function below 100 km has seasonal dependence closely related to mixed-layer depth, fluctuating between $k^{-1}$ and $k^{-2}$. 

In the remainder of this article, we account for these discrepancies by generalizing the uniformly stratified surface quasigeostrophic model \citep{held_surface_1995} to account for variable stratification (section \ref{S-inversion}). Generally, we find that the surface kinetic energy spectrum in surface quasigeostrophic turbulence depends on the stratification's vertical structure (section \ref{S-turbulence}); we recover the \cite{blumen_uniform_1978} spectral predictions only in the limit of uniform stratification. Stratification controls the kinetic energy spectrum by modifying the interaction range of surface quasigeostrophic eddies, and we illustrate this dependence by examining the turbulence under various idealized stratification profiles (section \ref{S-idealized}). We then apply the theory to the North Atlantic in both winter and summer, and find that the surface transfer function is seasonal, with a $\mathcal{F}(k) \sim k^{-3/2}$ dependence in winter and a $\mathcal{F}(k) \sim k^{-1/2}$ dependence in summer. Moreover, in the wintertime North Atlantic, the theory predicts a surface kinetic energy spectrum between $k^{-4/3}$ and $k^{-7/3}$, which is consistent with both observations and numerical simulations (section \ref{S-ECCO}). Finally, in section 6, we discuss the validity of theory at other times and locations.

\section{The inversion function} \label{S-inversion}

\subsection{Physical space equations}

Consider an ocean of depth $H$ with zero interior potential vorticity $(q=0)$ so that the geostrophic streamfunction satisfies 
\begin{equation}\label{eq:zero_pv}
	\lap \psi  + \pd{}{z}\left(\frac{1}{\sigma^2}\pd{\psi}{z}\right) = 0 \quad \text{for } z\in(-H,0).
\end{equation}
In this equation, $\lap$ is the horizontal Laplacian, $\psi$ is the geostrophic streamfunction, and 
\begin{equation}
	\sigma(z) = N(z)/f,
\end{equation}
where $N(z)$ is the depth-dependent buoyancy frequency and $f$ is the constant local value of the Coriolis frequency. We refer to $\sigma(z)$ as the \emph{stratification} for the remainder of this article. The horizontal geostrophic velocity is obtained from $\vec u = \unit z \times \grad \vec \psi$ where $\unit z$ is the vertical unit vector.

The upper surface potential vorticity is given by \citep{bretherton_critical_1966}
\begin{equation}\label{eq:theta_def}
	\theta = - \frac{1}{\sigma_0^2} \pd{\psi}{z}\bigg|_{z=0},
\end{equation}
where $\sigma_0 = \sigma(0)$. The surface potential vorticity is related to the surface buoyancy anomaly through 
\begin{equation}
	b|_{z=0} = -f\,\sigma_0^2\,\theta.
\end{equation}
The time-evolution equation at the upper boundary is given by
\begin{equation}\label{eq:theta_equation}
	\pd{\theta}{t} + \J{\psi}{\theta} = F-D \quad \text{at } z=0,
\end{equation}
where $\J{\theta}{\psi} = \partial_x \theta \, \partial_y \psi - \partial_y \theta \, \partial_x \psi$ represents the advection of $\theta$ by the horizontal geostrophic velocity $\vec u$, $F$ is the buoyancy forcing at the upper boundary, and $D$ is the dissipation. 

We assume a bottom boundary condition of
\begin{equation}
	\psi \rightarrow 0 \text{ as } z\rightarrow-\infty,
\end{equation}
which is equivalent to assuming the bottom boundary, $z=-H$, is irrelevant to the dynamics. In section \ref{S-ECCO}, we find that this assumption is valid in the mid-latitude North Atlantic open ocean at horizontal scales smaller than $\approx 500$ km. We consider alternative boundary conditions in appendix A.

\subsection{Fourier space equations}

Assuming a doubly periodic domain in the horizontal prompts us to consider the Fourier expansion of $\psi$,
\begin{equation}\label{eq:fourier}
	\psi(\vec r, z,t) = \sum_{\vec k} \hat \psi_{\vec k}(t) \, \Psi_{k}(z) \, \e^{\irm \vec k \cdot \vec r},
\end{equation}
where $\vec r = (x,y)$ is the horizontal position vector, $\vec k = (k_x, k_y)$ is the horizontal wavevector,  and $k=\abs{\vec k}$ is the horizontal wavenumber. The wavenumber dependent non-dimensional vertical structure, $\Psi_{k}(z)$, is determined by the boundary-value problem\footnote{To derive the vertical structure equation \eqref{eq:Psi_equation}, substitute the Fourier representation \eqref{eq:fourier} into the vanishing potential vorticity condition \eqref{eq:zero_pv}, multiply by $\mathrm{e}^{-i\vec l \cdot \vec r}$, take an area average, and use the identity $$ \frac{1}{A} \, \int_{A} \mathrm{e}^{\irm \left(\vec k-\vec l\right)\cdot \vec r} \, \mathrm{d} {\vec r} =  \delta_{\vec k,\vec l} $$ where $\delta_{\vec k, \vec l}$ is the Kronecker delta, and $A$ is the horizontal area of the periodic domain.} 
\begin{equation}\label{eq:Psi_equation}
	- \d{}{z} \left(\frac{1}{\sigma^2} \d{\Psi_k}{z} \right) + k^2 \, \Psi_k = 0,
\end{equation}
with upper boundary condition
\begin{equation}\label{eq:Psi_upper}
	\Psi_k(0) = 1
\end{equation}
and bottom boundary condition 
\begin{equation}\label{eq:Psi_lower}
	\Psi_k \rightarrow 0 \text{ as } z\rightarrow-\infty.
\end{equation}
The upper boundary condition \eqref{eq:Psi_upper} is a normalization for the vertical structure, $\Psi_k(z)$, chosen so that
\begin{equation}\label{eq:fourier_zero}
	\psi(\vec r, z=0,t) = \sum_{\vec k} \hat \psi_{\vec k}(t) \, \e^{\irm \vec k \cdot \vec r}.
\end{equation}
Consequently, the surface potential vorticity \eqref{eq:theta_def} is given by
\begin{equation}\label{eq:theta_fourier}
	\theta(\vec r, t) = \sum_{\vec k} \hat \theta_{\vec k}(t) \, \e^{\irm \vec k \cdot \vec r},
\end{equation}
where
\begin{equation}\label{eq:theta_inversion}
	\thetak = - m(k) \, \psik,
\end{equation}
and the inversion function $m(k)$ (with dimensions of inverse length) is defined as
\begin{equation}\label{eq:mk}
	m(k)= \frac{1}{\sigma_0^2} \, \d{\Psi_k(0)}{z}.
\end{equation}
In all our applications, we find the inversion function to be a positive monotonically increasing function of $k$ [i.e., $m(k)>0$ and $\mathrm{d}m/\mathrm{d}k\geq0$]. 
The inversion function is related to the transfer function \eqref{eq:transfer_func} through 
\begin{equation}\label{eq:transfer_inversion}
	\mathcal{F}(k) = \frac{1}{f \, \sigma_0^2 \, m(k)} =  \left[ f \, \d{\Psi_k(0)}{z} \right]^{-1},
\end{equation}
which shows that the transfer function, evaluated at a wavenumber $k$, is related to the characteristic vertical scale of $\Psi_k(z)$.


\subsection{The case of constant stratification}\label{SS-constant_strat}

To recover the well-known case of the uniformly stratified surface quasigeostrophic model \citep{held_surface_1995}, set $\sigma = \sigma_0$. Then  solving the vertical structure equation \eqref{eq:Psi_equation} along with boundary conditions \eqref{eq:Psi_upper} and \eqref{eq:Psi_lower} yields the exponentially decaying vertical structure, 
\begin{equation}
	\Psi_k(z) = \e^{\sigma_0 \, k \, z}.
\end{equation}
Substituting $\Psi_k(z)$ into the definition of the inversion function \eqref{eq:mk}, we obtain
 \begin{equation}
	 m(k) = k/\sigma_0,
\end{equation}
and hence [through the inversion relation \eqref{eq:theta_fourier}] a linear-in-wavenumber inversion relation of 
\begin{equation}
	\hat \theta_{\vec k} = -(k/\sigma_0) \, \hat \psi_{\vec k}.
\end{equation}

In appendix A, we show that $m(k) \rightarrow k/\sigma_0$ as $k\rightarrow \infty$ for arbitrary stratification $\sigma(z)$. Therefore, at sufficiently small horizontal scales, surface quasigeostrophic dynamics behaves as in constant stratification regardless of the functional form of $\sigma(z)$.

\section{Surface quasigeostrophic turbulence}\label{S-turbulence} 

Suppose a two-dimensional barotropic fluid is forced in the wavenumber interval $[k_1,k_2]$. In such a fluid, \cite{kraichnan_inertial_1967} argued that two inertial ranges will form: one inertial range for $k<k_1$ where kinetic energy cascades to larger scales and another inertial range for $k>k_2$ where enstrophy cascades to smaller scales. Kraichnan's argument depends on three properties of two-dimensional vorticity dynamics. First, that there are two independent conserved quantities; namely, the kinetic energy and the enstrophy. Second, that turbulence is sufficiently local in wavenumber space   so that the only available length scale is $k^{-1}$. Third, that the inversion relation between the vorticity and the streamfunction is scale invariant. 

There are two independent conserved quantities in surface quasigeostrophic dynamics, as in Kraichnan's two-dimensional fluid; namely the total energy, $E$, and the surface potential enstrophy, $P$. 
However, the second and third properties of two-dimensional vorticity dynamics do not hold for surface quasigeostrophic dynamics. Even if the turbulence is local in wavenumber space, there are two available length scales at each wavenumber $k$; namely, $k^{-1}$ and $[m(k)]^{-1}$. Moreover, the inversion relation \eqref{eq:theta_inversion} is generally not scale invariant.\footnote{A function $m(k)$ is scale invariant if $m(\lambda k) = \lambda^s m(k)$ for all $\lambda$, where $s$ is a real number. In particular, note that power laws, $m(k) = k^\alpha$, are scale invariant.} Therefore, the arguments in \cite{kraichnan_inertial_1967} do not hold in general for surface quasigeostrophic dynamics.

Even so, in the remainder of this section we show that there is a net inverse cascade of total energy and a net forward cascade of surface potential enstrophy even if there are no inertial ranges in the turbulence. Then we consider the circumstances under which we expect inertial ranges to form. Finally, assuming the existence of an inertial range, we derive the spectra for the cascading quantities. We begin, however, with some definitions.

\subsection{Quadratic quantities}
The two quadratic quantities needed for the cascade argument are the volume-integrated total mechanical energy per mass per unit area, 
\begin{equation}\label{eq:total_energy}
\begin{aligned}
	E &= \frac{1}{2\,A} \int_V \left( \abs{\grad \psi}^2 + \frac{1}{\sigma^{2}} \abs{\pd{\psi}{z}}^2 \right)\mathrm{d}V \\
	  &=  - \frac{1}{2}\,\overline{\psi|_{z=0} \, \theta} = \frac{1}{2} \sum_{\vec k} m(k) \, \abs{\hat \psi_{\vec k}}^2,
\end{aligned}
\end{equation}
and the upper surface potential enstrophy, 
\begin{equation}\label{eq:potential_enstrophy}
	P = \frac{1}{2} \, \overline{\theta^2} = \frac{1}{2} \sum_{\vec k} \left[m(k)\right]^2 \abs{\hat \psi_{\vec k}}^2,
\end{equation}
where the overline denotes an area average over the periodic domain.
Both quantities are time-independent in the absence of forcing and dissipation, as can be seen by multiplying the time-evolution equation \eqref{eq:theta_equation} by either $-\psi|_{z=0}$ or $\theta$ and taking an area average.




Two other quadratics we use are the surface kinetic energy
\begin{equation}\label{eq:kinetic_energy}
	K = \frac{1}{2} \overline{\abs{\grad \psi}_{z=0}^2} = \frac{1}{2} \sum_{\vec k} k^2 \, \abs{\hat \psi_{\vec k}}^2
\end{equation}
and the surface streamfunction variance
\begin{equation}\label{eq:stream_var}
	S = \frac{1}{2} \overline{\left(\psi|_{z=0}\right)^2} = \frac{1}{2} \sum_{\vec k} \abs{\hat \psi_{\vec k}}^2.
\end{equation}
Moreover, the isotropic spectrum $\mathscr{A}(k)$ of a quadratic quantity $A$ is defined by
\begin{equation}
	A = \int_{0}^\infty \mathscr{A}(k) \, \di k,
\end{equation}
so that the isotropic spectra of $E, P, K,$ and $S$ are given by $\Esc(k), \Psc(k), \Ksc(k),$ and $\Ssc(k)$. The isotropic spectra are then related by
\begin{equation}\label{eq:variance_relations1}
	\Psc(k) = m(k) \, \Esc(k) = \left[m(k)\right]^2 \, \Ssc(k)
\end{equation}
and 
\begin{equation}\label{eq:variance_relations2}
	\Ksc(k) = k^2 \, \Ssc(k).
\end{equation}

For $\mathscr{A}(k) = \Esc(k)$ or $\mathscr{A}(k)=\Psc(k)$, there is a time-evolution equation of the form \citep{gkioulekas_new_2007}
\begin{equation}\label{eq:spectral_transfer}
	\pd{\mathscr{A}(k)}{t} + \pd{\Pi_A(k)}{k} =F_A(k) - D_A(k),
\end{equation}
where $\Pi_A(k)$ is the transfer of the spectral density $\mathscr{A}(k)$ from $(0,k)$ to $(k,\infty)$, and $D_A(k)$ and $F_A(k)$ are the dissipation and forcing spectra of $A$, respectively. In an inertial range where $A$ is the cascading quantity, then $\Pi_A(k) = \varepsilon_A$ where $\varepsilon_A$ is a constant and thus $\partial \Pi_A(k)/\partial k = 0$.

\subsection{The inverse and forward cascade}

For a fluid with the variable stratification inversion relation \eqref{eq:theta_inversion} that is forced in the wavenumber interval $[k_1,k_2]$, \cite{gkioulekas_new_2007} prove the following two inequalities for stationary turbulence,
\begin{align}
	\label{eq:gk_1}
	\int_{0}^k \d{m(k')}{k'} \,\Pi_{E}(k') \, \di{k'} < 0, \, \, \text{for all } k>k_2,\\
	\label{eq:gk_2}
	\int_{k}^\infty \d{m(k')}{k'} \, \frac{\Pi_P(k')}{[m(k')]^2} \,\di{k'} > 0, \, \, \text{for all } k<k_1.
\end{align}
These two inequalities do not require the existence of inertial ranges, only that the inversion function $m(k)$ is an increasing function of $k$.
 Therefore, if $\mathrm{d}m(k)/\mathrm{d}k>0$, then there is a net inverse cascade of total energy and a net forward cascade of surface potential enstrophy.

\subsection{When do inertial ranges form?}

The lack of scale invariance along with the presence of two length scales, $k^{-1}$ and $[m(k)]^{-1}$, prevents the use of the \cite{kraichnan_inertial_1967} argument to establish the existence of an inertial range. However, suppose that
in a wavenumber interval, $[k_a,k_b]$, the inversion function takes the power law form
\begin{equation}\label{eq:mk_power}
	m(k) \approx m_\alpha \, k^{\alpha}, 
\end{equation}
where $m_\alpha > 0$ and $\alpha >0$.
Then, in this wavenumber interval, the inversion relation takes the form of the $\alpha$-turbulence inversion relation \citep{pierrehumbert_spectra_1994}, 
\begin{equation}\label{eq:alpha_inv}
		\hat \xi_{\vec k} = - k^{\alpha} \, \hat \psi_{\vec k},
\end{equation}
with $\xi = \theta/m_\alpha$.
The inversion relation \eqref{eq:alpha_inv} is then scale invariant in the wavenumber interval $[k_a,k_b]$. Moreover, $k^{-1}$ is the only available length scale if the turbulence is sufficiently local in wavenumber space. It follows that if the wavenumber interval $[k_a,k_b]$ is sufficiently wide (i.e., $k_a \ll k_b$), then Kraichnan's argument applies to the turbulence over this wavenumber interval and inertial ranges are expected to form. 




\subsection{The \cite{tulloch_theory_2006} argument}

If we assume the existence of inertial ranges, then we can adapt the cascade argument of \cite{tulloch_theory_2006} to general surface quasigeostrophic fluids to obtain predictions for the cascade spectra.

In the inverse cascade inertial range, we must have $\Pi_E(k)= \varepsilon_E$ where  $\varepsilon_E$ is a constant. Assuming locality in wavenumber space, we have
\begin{equation}\label{eq:epsilon_E}
	\varepsilon_E \sim  \frac{k \, \Esc(k)}{\tau(k)},
\end{equation}
where $\tau(k)$ is a spectrally local timescale\footnote{A spectrally local timescale is appropriate so long as $m(k)$ grows less quickly than $k^2$. Otherwise, a non-local timescale must be used \citep{kraichnan_1971,watanabe_unified_2004}.}. If we further assume that the timescale $\tau(k)$ is determined by the kinetic energy spectrum, $\Ksc(k)$, then dimensional consistency requires 
\begin{equation}\label{eq:turbulent_time}
	\tau(k) \sim \left[k^3 \, \Ksc(k) \right]^{-1/2}.
\end{equation}
Substituting this timescale into equation \eqref{eq:epsilon_E} and using the relationship between the energy spectrum, $\Esc(k)$, and the streamfunction variance spectrum, $\Ssc(k)$, in equations \eqref{eq:variance_relations1} and  \eqref{eq:variance_relations2}, we obtain the total energy spectrum in the inverse cascade inertial range,
\begin{equation}\label{eq:Espec_inv}
	\Esc(k) \sim \varepsilon_E^{2/3} \, k^{-7/3} \, \left[m(k)\right]^{1/3}.
\end{equation}
Analogously, in the forward cascade inertial range, we must have $\Pi_P(k) = \varepsilon_P$ where $\varepsilon_P$ is a constant. 
A similar argument yields the surface potential enstrophy spectrum in the forward cascade inertial range,
\begin{equation}\label{eq:Pspec_forward}
	\Psc(k) \sim \varepsilon_P^{2/3} \, k^{-7/3} \, \left[m(k)\right]^{2/3}.
\end{equation}


The predicted spectra \eqref{eq:Espec_inv} and \eqref{eq:Pspec_forward} are not uniquely determined by dimensional analysis. Rather than assuming that the spectrally local timescale $\tau(k)$ is determined by the kinetic energy spectrum, $\Ksc(k)$, we can assume that $\tau(k)$ is determined by the total energy spectrum, $\Esc(k)$, or the surface potential enstrophy spectrum, $\Psc(k)$.\footnote{These assumptions lead to timescales of  $\tau(k) \sim \left[k^4 \, \Esc(k)\right]^{-1/2}$ and $\tau(k) \sim \left[k^3 \, \Psc(k)\right]^{-1/2}$, respectively.}
Either choice will result in cascade spectra distinct from \eqref{eq:Espec_inv} and \eqref{eq:Pspec_forward}. However, by assuming that the timescale $\tau(k)$ is determined by the kinetic energy spectrum, the resulting cascade spectra agree with the $\alpha$-turbulence predictions of \cite{pierrehumbert_spectra_1994} when the inversion function takes the power law form \eqref{eq:mk_power}.

For later reference, we provide the expressions for the inverse and forward cascade surface kinetic energy spectra.  Using either the inverse cascade spectrum \eqref{eq:Espec_inv} or forward cascade spectrum \eqref{eq:Pspec_forward} along with the relations between the various spectra [equations \eqref{eq:variance_relations1} and \eqref{eq:variance_relations2}], we obtain 
	\begin{equation}\label{eq:inverse_KE}
		\Ksc(k) \sim \varepsilon_E^{2/3} \, k^{-1/3} \, \left[m(k)\right]^{-2/3}
	\end{equation}
	in the inverse cascade and 
	\begin{equation}\label{eq:forward_KE}
			\Ksc(k) \sim \varepsilon_P^{2/3} \, k^{-1/3} \, \left[m(k)\right]^{-4/3}
	\end{equation}
	in the forward cascade. 

Finally, we note that the vorticity spectrum,
	\begin{equation}
		\mathscr{Z}(k) = k^2 \, \mathscr{K}(k),
	\end{equation}
	is an increasing function of $k$ if $m(k)$ is flatter than $k^{5/4}$. In particular, at small scales, we expect $m(k) \sim k$ (section \ref{S-inversion}\ref{SS-constant_strat}), implying a vorticity spectrum of $\mathscr{Z}(k) \sim k^{1/3}$. Such an increasing vorticity spectrum implies high Rossby numbers and the breakdown of geostrophic balance at small scales.

\section{Idealized stratification profiles}\label{S-idealized}

In this section we provide analytical solutions for $m(k)$ in the cases of an increasing and decreasing piecewise constant stratification profiles as well as in the case of exponential stratification. These idealized stratification profiles then provide intuition for the inversion function's functional form in the case of an arbitrary stratification profile, $\sigma(z)$.

\subsection{Piecewise constant stratification}

	\begin{figure*}
	\centerline{\includegraphics[width=33pc]{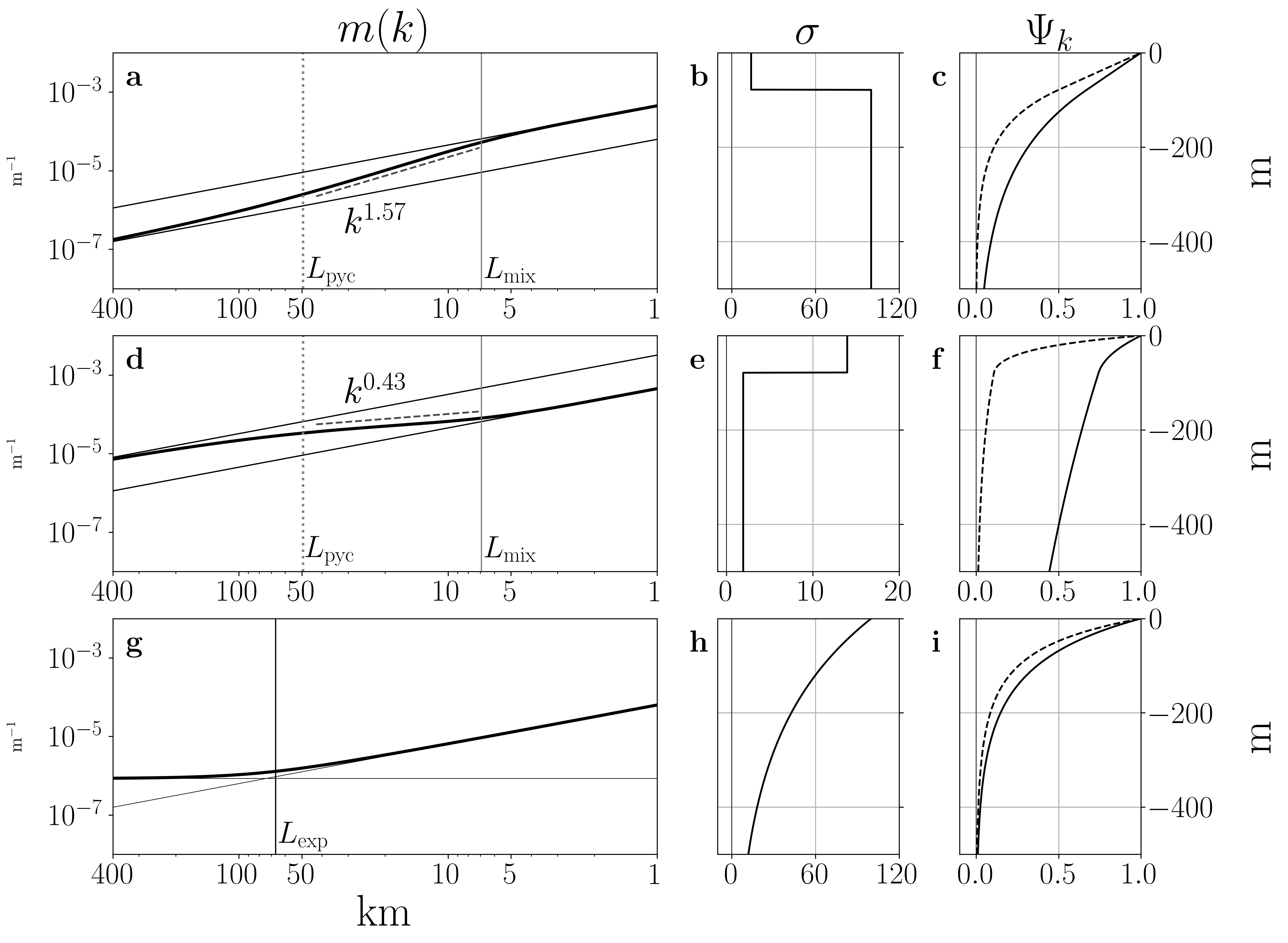}}
	\caption{Log-log plots of the inversion function, $m(k)$ [panels (a), (d), and (g)], for three stratification profiles [panels (b), (e), and (h)] and the resulting streamfunctions at the two horizontal length scales of 50 km (dashed) and 100 km (solid) [for panels (c) and (i)] or 2 km and 10 km [panel (f)]. In the first two inversion function plots [panels (a) and (d)], the thin solid diagonal lines represent the two linear asymptotic states of $k/\sigma_0$ and $k/\sigma_\mathrm{pyc}$. The vertical solid line is the mixed-layer length scale $L_\mathrm{mix}$, given by equation \eqref{eq:mixed_length}, whereas the vertical dotted line is the pycnocline length scale $L_\mathrm{pyc}$, given by equation \eqref{eq:therm_length}. The power $\alpha$, where $m(k)/k^{\alpha} \approx \mathrm{constant}$, is computed by fitting a straight line to the log-log plot of $m(k)$ between $2\pi/L_\mathrm{mix}$ and $2\pi/L_\mathrm{pyc}$. This straight line is shown as a grey dashed line in panels (a) and (d). In panel (g), the thin  diagonal line is the linear small-scale limit, $m(k)\approx k/\sigma_0$, whereas the thin horizontal line is the constant large-scale limit, $m(k) = 2/(\sigma_0^2 \, h_\mathrm{exp})$. Finally, the solid vertical lines in panel (g) indicate the horizontal length scale $L_\mathrm{exp} = 2\pi/k_\mathrm{exp}$ [equation \eqref{eq:kexp}] induced by the exponential stratification. Further details on the stratification profiles are in the text.}
	\label{F-mk_constlinconst}
  	\end{figure*}

	Consider the piecewise constant stratification profile, given by
		\begin{equation}\label{eq:step_strat}
		\sigma(z) = 
		\begin{cases}
			\sigma_0  \, &\text{for } -h < z \leq 0 \\
			\sigma_\mathrm{pyc} \, &\text{for } \infty < z \leq -h.
		\end{cases}
	\end{equation}
	This stratification profile consists of an upper layer of thickness $h$ with constant stratification $\sigma_0$ overlying an infinitely deep layer with constant stratification $\sigma_\mathrm{pyc}$.  If $\sigma_0 < \sigma_\mathrm{pyc}$, then this stratification profile is an idealization of a weakly stratified mixed-layer overlying an ocean of stronger stratification. 
     See panels (b) and (e) in figure \ref{F-mk_constlinconst} for an illustration.  
	
	
	For this stratification profile, an analytical solution is possible, with the solution provided in appendix B. The resulting inversion function is
	\begin{equation}
		m(k) = 
		\frac{k}{\sigma_0} \left[ \frac{\cosh\left(\sigma_0 h k\right) + \left(\frac{\sigma_\mathrm{pyc}}{\sigma_0}\right) \sinh\left(\sigma_0  h k\right) }{\sinh\left(\sigma_0  hk\right) + \left(\frac{\sigma_\mathrm{pyc}}{\sigma_0}\right) \cosh\left(\sigma_0  h k\right) } \right].
	\end{equation}
	At small horizontal scales, $2\pi/k \ll L_\mathrm{mix}$, where 
	\begin{equation}\label{eq:mixed_length}
		L_\mathrm{mix} = 2\,\pi \, \sigma_0 \, h,
	\end{equation}
	the inversion function takes the form $m(k) \approx k/\sigma_0$, as expected from the uniformly stratified theory \citep{held_surface_1995}. At large horizontal scales, $2\pi/k \gg L_\mathrm{pyc}$, where 
	\begin{equation}\label{eq:therm_length}
	 L_\mathrm{pyc} =  2 \, \pi 
	 \begin{cases}
	 		  \, \sigma_\mathrm{pyc} \, h  \, &\text{if } \sigma_0 \leq \sigma_\mathrm{pyc} \\
	 		\sigma_\mathrm{0}^2 \, h /\sigma_\mathrm{pyc}  \, &\text{if } \sigma_0 > \sigma_\mathrm{pyc},
	 \end{cases}
	\end{equation}
	then the inversion function takes the form $m(k) \approx k/\sigma_\mathrm{pyc}$, because at large horizontal scales, the ocean will seem to have constant stratification $\sigma_\mathrm{pyc}$.
	
	The functional form of the inversion function at horizontal scales between $L_\mathrm{mix}$ and $L_\mathrm{pyc}$ depends on whether $\sigma(z)$ is a decreasing or increasing function. If $\sigma(z)$ is a decreasing function, with $\sigma_0 < \sigma_\mathrm{pyc}$, then we obtain a mixed-layer like stratification profile and the inversion function steepens to a super linear wavenumber dependence at these scales. An example is shown in figure \ref{F-mk_constlinconst}(a)-(b). Here, the stratification abruptly jumps from a value of $\sigma_0 \approx 14$ to $\sigma_\mathrm{pyc} = 100$ at $z\approx-79$ m. Consequently, the inversion function takes the form $m(k) \sim k^{1.57}$ between $2\pi/L_\mathrm{pyc}$ and $2\pi/L_\mathrm{mix}$.
	In contrast, if $\sigma_0 > \sigma_\mathrm{pyc}$ then the inversion function flattens to a sublinear wavenumber dependence for horizontal scales between $L_\mathrm{mix}$ and $L_\mathrm{pyc}$. An example is shown in figure \ref{F-mk_constlinconst}(d)-(e), where the stratification abruptly jumps from $\sigma_0 \approx 14$ to $\sigma_\mathrm{pyc} \approx 2$ at $z\approx-79$ m. In this case, the inversion function has a sublinear wavenumber dependence, $m(k) \sim k^{0.43}$, between $2\pi/L_\mathrm{pyc}$ and $2\pi/L_\mathrm{mix}$.
		
	\begin{figure*}
	\centerline{\includegraphics[width=33pc]{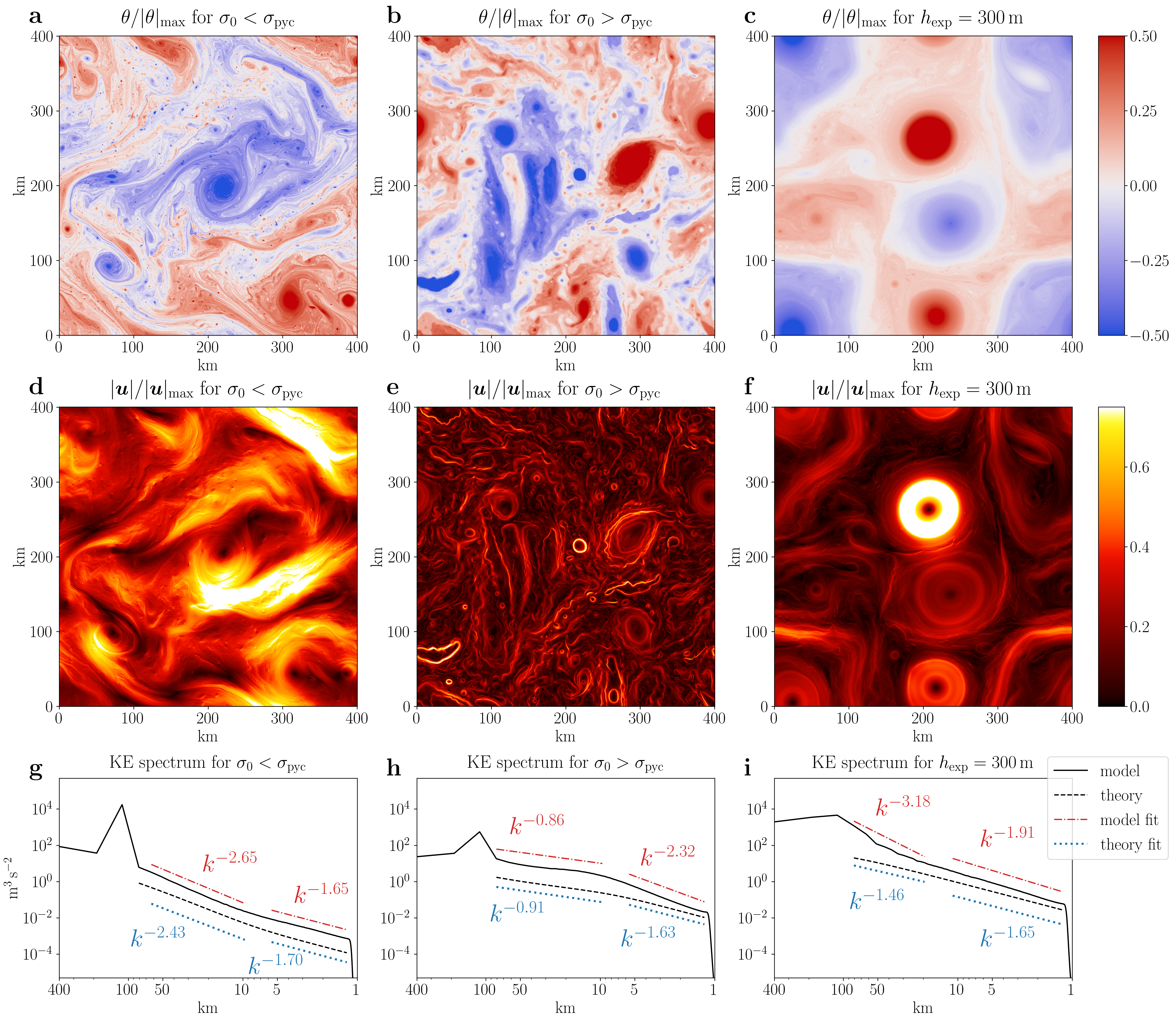}}
	\caption{Results of three pseudo-spectral simulations, forced at approximately 100 km, with $1024^2$ horizontal grid points. See appendix C for a description of the numerical model. The first simulation [panels (a), (d), and (g)] corresponds to the stratification profile and inversion function shown in figure \ref{F-mk_constlinconst}(a)-(b), the second simulations [panels (b), (e), and (h)] corresponds to the stratification profile and inversion function shown in figure \ref{F-mk_constlinconst}(d)-(e), and the third simulation corresponds to the stratification profile and inversion function shown in figure \ref{F-mk_constlinconst}(g)-(h). Plots (a), (b), and (c) are snapshots of the surface potential vorticity, $\theta$, normalized by its maximum value in the snapshot. Plots (d), (e), and (f) are snapshots of the horizontal speed $\abs{\vec u }$ normalized by its maximum value in the snapshot. Plots (g), (h), and (i) show the model kinetic energy spectrum (solid black line) along with the prediction given by equation \eqref{eq:forward_KE} (dashed black line). We also provide linear fits to the model kinetic energy spectrum (dash-dotted red line) and to the predicted spectrum (dotted blue line).}
	\label{F-model_runs_mixed}
	\end{figure*}

	By fitting a power law, $k^\alpha$, to the inversion function, we do not mean to imply that $m(k)$ indeed takes the form of a power law. Instead, the purpose of obtaining the estimated power $\alpha$ is to apply the intuition gained from $\alpha$-turbulence \citep{pierrehumbert_spectra_1994,smith_turbulent_2002,sukhatme_local_2009,burgess_kraichnanleithbatchelor_2015,foussard_relative_2017}  to surface quasigeostrophic turbulence. In $\alpha$-turbulence, an active scalar $\xi$, defined by the power law inversion relation \eqref{eq:alpha_inv}, is materially conserved in the absence of forcing and dissipation [that is, $\xi$ satisfies the time-evolution equation \eqref{eq:theta_equation} with $\theta$ replaced by $\xi$]. The scalar $\xi$ can be thought of as a generalized vorticity; if $\alpha=2$ we recover the vorticity of two-dimensional barotropic model. If $\alpha=1$, $\xi$ is proportional to the surface buoyancy in the uniformly stratified surface quasigeostrophic model. To discern how $\alpha$ modifies the dynamics, we consider a point vortex $\xi \sim \delta(r)$, where $r$ is the horizontal distance from the vortex and $\delta(r)$ is the Dirac delta. If $\alpha=2$, we obtain  $\psi(r) \sim \log(r)/2\pi$; otherwise, if $0 < \alpha < 2$, we obtain $\psi(r) \sim - C_\alpha/r^{2-\alpha}$ where $C_\alpha>0$ \citep{iwayama_greens_2010}. 
	Therefore, larger $\alpha$ leads to vortices with a longer interaction range whereas smaller $\alpha$ leads to a shorter interaction range. 
	
	More generally, $\alpha$ controls the spatial locality of the resulting turbulence. In two-dimensional turbulence ($\alpha = 2$), vortices induce flows reaching far from the vortex core and the combined contributions of distant vortices dominates the local fluid velocity. These flows are characterized by thin filamentary $\xi$-structures due to the dominance of large-scale strain \citep{watanabe_unified_2004}.
		As we decrease $\alpha$, the turbulence becomes more spatially local, the dominance of large-scale strain weakens, and a secondary instability becomes possible in which filaments roll-up into small vortices; the resulting turbulence is distinguished by vortices spanning a wide range of horizontal scales, as in uniform stratification surface quasigeostrophic turbulence \citep{pierrehumbert_spectra_1994, held_surface_1995}. As $\alpha$ is decreased further the $\xi$ field becomes spatially diffuse because the induced velocity, which now has small spatial scales, is more effective at mixing small-scale inhomogeneities in $\xi$ \citep{sukhatme_local_2009}.	  		
	
	These expectations are confirmed in the simulations shown in figure \ref{F-model_runs_mixed}.
	The simulations are set in a doubly periodic square with side length 400 km and are forced at a horizontal scale of 100 km. Large-scale dissipation is achieved through a linear surface buoyancy damping whereas an exponential filter is applied at small scales. In the case of a mixed-layer like stratification, with $\sigma_0 < \sigma_\mathrm{pyc}$, the $\theta$ field exhibits thin filamentary structures (characteristic of the $\alpha=2$ case) as well as vortices spanning a wide range of horizontal scales (characteristic of the $\alpha=1$ case). In contrast, although the $\sigma_0 > \sigma_\mathrm{pyc}$ simulation exhibits vortices spanning a wide range of scales, no large-scale filaments are evident. Instead, we see that the surface potential vorticity is spatially diffuse. These contrasting features are consequences of the induced horizontal velocity field. The mixed-layer like case has a velocity field dominated by large-scale strain, which is effective at producing thin filamentary structures. In contrast the velocity field in the $\sigma_0> \sigma_\mathrm{pyc}$ case consists of narrow meandering currents, which are effective at mixing away small-scale inhomogeneities. 
	

	Both the predicted [equation \eqref{eq:forward_KE}] and diagnosed surface kinetic energy spectra are plotted in figure \ref{F-model_runs_mixed}. In the mixed-layer like case, with $\sigma_0 < \sigma_\mathrm{pyc}$, the predicted and diagnosed spectrum are close, although the diagnosed spectrum is steeper at large scales \citep[a too steep spectrum is also observed in the $\alpha=1$ and $\alpha=2$ cases, see][]{schorghofer_energy_2000}. In the $\sigma_0 > \sigma_\mathrm{pyc}$ case, the large-scale spectrum agrees with the predicted spectrum. However, at smaller scales, the model spectrum is significantly steeper.
	
	\begin{figure*}
	\centerline{\includegraphics[width=33pc]{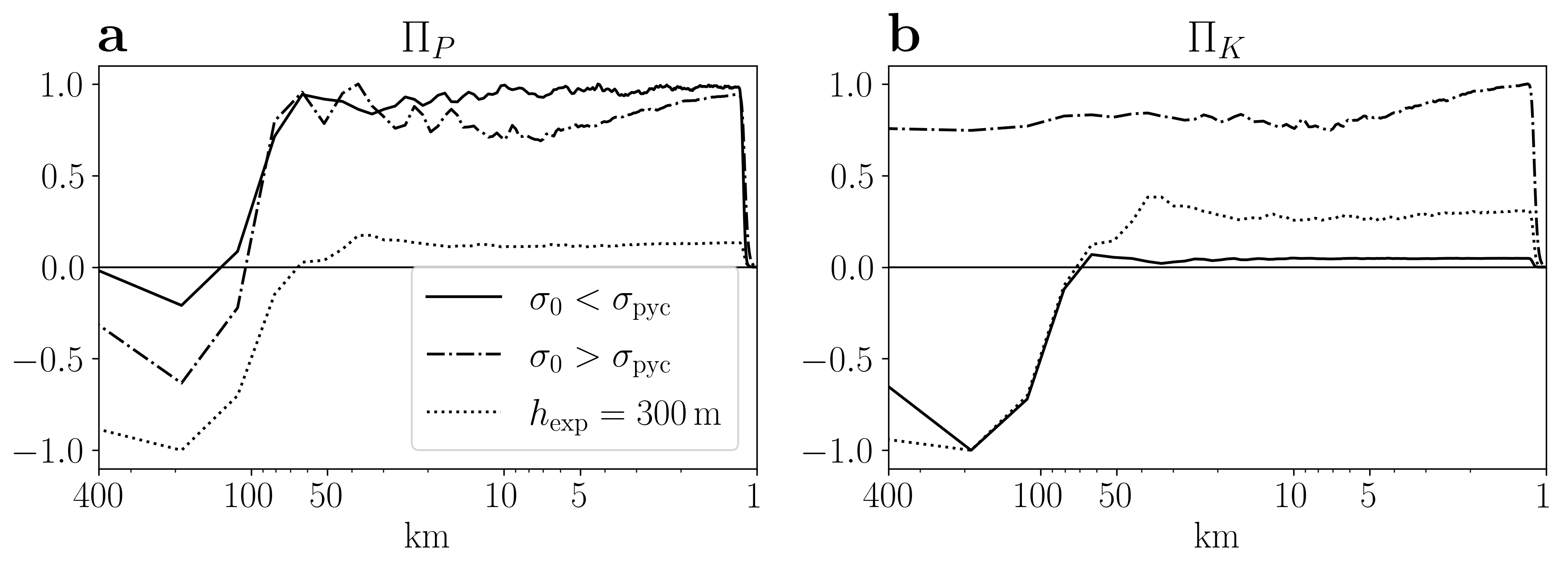}}
	\caption{The spectral density transfer functions for surface potential enstrophy (a) and surface kinetic energy (b) normalized by their absolute maximum for the three simulations in figure \ref{F-model_runs_mixed}.}
	\label{F-model_runs_transfer}
	\end{figure*}
	
	The derivation of the predicted spectra in section \ref{S-turbulence} assumed the existence of an inertial range, which in this case means $\Pi_P(k)=$ constant. To verify whether this assumption holds, we show in figure \ref{F-model_runs_transfer}(a) the spectral transfer of surface potential enstrophy for both the $\sigma_0<\sigma_\mathrm{pyc}$ and the $\sigma_0>\sigma_\mathrm{pyc}$ cases.
	In the mixed-layer like case, with $\sigma_0 < \sigma_\mathrm{pyc}$, an approximate inertial range forms with some deviations at larger scales. However, in the $\sigma_0 > \sigma_\mathrm{pyc}$ case, $\Pi_P$ is an increasing function at small scales, which indicates that the spectral density of surface potential enstrophy, $\Psc(k)$, is diverging at these scales. That is, at small scales, there is a depletion of $\Psc(k)$ and this depletion is causing the steepening of the kinetic energy spectrum at small scales in figure \ref{F-model_runs_mixed}.
	
	
	The steepening of the model spectrum at small scales may be a consequence of the inversion function being steeper at small scales than at large scales.
	We tested this hypothesis with two additional simulations having a prescribed inversion function of the form
	\begin{equation}
		m(k) =
		\begin{cases}
			k^{\alpha_1} \quad \text{for }  k<k_0 \\
			C \, k^{\alpha_2} \quad \text{for } k>k_0,
		\end{cases}
	\end{equation}
	where $\alpha_1$,$\alpha_2$, and $k_0$ are positive numbers, and $C$ is chosen to ensure that $m(k)$ is continuous. In the first simulation, we chose $\alpha_1 = 3/2$ and $\alpha_2=1/2$ so that $m(k)$ is flatter at small scales than at large scales (as in the mixed-layer like case). We obtained an approximate inertial range and the model spectrum is close to the predicted surface kinetic energy spectrum \eqref{eq:forward_KE}. In the second simulation, we chose $\alpha_1 = 1/2$ and $\alpha_2 = 3/2$ so that $m(k)$ is steeper at small scales than at large scales (as in the $\sigma_0>\sigma_\mathrm{pyc}$ case). We found that $\Pi_P(k)$ is an increasing function (so that no inertial range exists) and we obtained a model surface kinetic energy spectrum that is much steeper than predicted. It is not clear why the inertial range theory fails in this case, and the failure may be a consequence of the finite model resolution.
					
	\subsection{An exponentially stratified ocean}\label{SS-exp}
	
	Now consider the exponential stratification profile
	\begin{equation}\label{eq:sigma_exp}
		\sigma = \sigma_0 \, \mathrm{e}^{z/h_\mathrm{exp}}.
	\end{equation}
	Substituting the stratification profile \eqref{eq:sigma_exp} into the vertical structure equation \eqref{eq:Psi_equation} with boundary conditions \eqref{eq:Psi_upper} and \eqref{eq:Psi_lower} yields the vertical structure
	\begin{equation}\label{eq:Psi_exp}
		\Psi_k(z) = \mathrm{e}^{z/h_\mathrm{exp}} \, \frac{ I_1\left(\mathrm{e}^{z/h_\mathrm{exp}} \sigma_0 \, h_\mathrm{exp} \, k \right) }{I_1\left(\sigma_0 \, h_\mathrm{exp} \, k \right)},
	\end{equation}
	where $I_n(z)$ is the modified Bessel function of the first kind of order $n$.
	
	To obtain the inversion function, we substitute the vertical structure \eqref{eq:Psi_exp} into the definition of the inversion function \eqref{eq:mk} to obtain
	\begin{equation}\label{eq:mk_exp}
	\begin{aligned}
		m(k) = &\frac{1}{\sigma_0^2 h_\mathrm{exp}} \, + \\ \qquad & \frac{k}{2\sigma_0} \left[  \frac{ I_0\left( \sigma_0 h_\mathrm{exp}  k \right) }{I_1\left(\sigma_0  h_\mathrm{exp} k \right)} + \frac{ I_2\left( \sigma_0  h_\mathrm{exp}  k \right) }{I_1\left(\sigma_0  h_\mathrm{exp} k \right)}  \right].
	\end{aligned}
	\end{equation}
	In the small-scale limit $k\gg 1/\left(\sigma_0\,h_\mathrm{exp}\right)$, the inversion function becomes $m(k) \approx k/\sigma_0$ as in constant stratification surface quasigeostrophic theory. In contrast, the large-scale limit $k\ll  1/\left(\sigma_0\,h_\mathrm{exp}\right)$ gives
	\begin{equation}\label{eq:mk_exp_large}
		m(k) \approx \frac{h_\mathrm{exp}}{4} \left( k_\mathrm{exp}^2 + k^2 \right),
	\end{equation}
	where $k_\mathrm{exp}$ is given by
	\begin{equation}\label{eq:kexp}
		k_\mathrm{exp} = \frac{2\, \sqrt{2}}{ \sigma_0 \, h_\mathrm{exp}}.
	\end{equation}
	As $k/k_\mathrm{exp} \rightarrow 0$, the inversion function asymptotes to a constant value and the vertical structure becomes independent of the horizontal scale $2\pi/k$, with $\Psi_k \rightarrow \Psi_0$ where
	\begin{equation}
	    \Psi_0(z) = \mathrm{e}^{2z/h_\mathrm{exp}}.
	\end{equation}
	Further increasing the horizontal scale no longer modifies $\Psi_k(z)$ and so vertical structure is arrested at $\Psi_0$.
 	
 	An example with $h_\mathrm{exp}=300$ m and $\sigma_0 = 100$ is shown in figure \ref{F-mk_constlinconst}(g)-(i). At horizontal scales smaller than $L_\mathrm{exp}=2\pi/k_\mathrm{exp}$, the inversion function rapidly transitions to the linear small-scale limit of $m(k)\approx k/\sigma_0$. In contrast, at horizontal scales larger than $L_\mathrm{exp}$, the large-scale approximation \eqref{eq:mk_exp_large} holds, and at sufficiently large horizontal scales, the inversion function asymptotes to constant value of $m(k) = h_\mathrm{exp}\,k_\mathrm{exp}^2/4$.
	
	The inversion relation implied by the inversion function \eqref{eq:mk_exp_large} is
	\begin{equation}\label{eq:theta_shallow}
		\hat \theta_{\vec k} \approx - \frac{h_\mathrm{exp}}{4} \left( k_\mathrm{exp}^2 + k^2 \right) \hat \psi_{\vec k},
	\end{equation}
	which is isomorphic to the inversion relation in the equivalent barotropic model \citep{larichev_weakly_1991}, with $k_\mathrm{exp}$ assuming the role of the deformation wavenumber. In this limit, the total energy and the surface potential enstrophy are not independent quantities to leading order in $k_\mathrm{exp}^2$.
	Using the relations between the various spectra [equations \eqref{eq:variance_relations1} and \eqref{eq:variance_relations2}] with an inversion function of the form $m(k)\approx m_0 +m_1 k^2$, we obtain $\Esc(k) \approx m_0 \, \Ssc(k) + m_1 \, \Ksc(k) $ and $\Psc(k) \approx m_0^2 \, \Ssc(k) + 2\, m_0 \, m_1 \Ksc(k)$; solving for $\Ssc(k)$ and $\Ksc(k)$ then yields
	\begin{equation}
		\Ssc(k) \approx \frac{2 \, m_0 \Esc(k) - \Psc(k)}{m_0^2},
	\end{equation}
	and
	\begin{equation}
		\Ksc(k) \approx  \frac{\Psc(k) - m_0 \, \Esc(k)}{m_0\,m_1},
	\end{equation}
	which are now the two independent quantities.
	Then, using an argument analogous to that in \cite{larichev_weakly_1991}, we find that 
	\begin{equation}
		\Ssc(k) \sim k^{-11/3}
	\end{equation}
	 in the inverse cascade inertial range whereas 
	 \begin{equation}\label{eq:equiv_KE}
	 	 \Ksc(k) \sim k^{-3}
	 \end{equation}
	 in the forward cascade inertial range.


	The implied dynamics are extremely local; a point vortex, $\theta(r) \sim \delta(r)$, leads to an exponentially decaying streamfunction, $\psi(r) \sim \exp(-k_\mathrm{exp}r)/\sqrt{k_\mathrm{exp}r}$ \citep{polvani_two-layer_1989}. Therefore, as for the $\sigma_0>\sigma_\mathrm{pyc}$ case above, we expect a spatially diffuse surface potential vorticity field and no large-scale strain. However, unlike the $\sigma_0>\sigma_\mathrm{pyc}$ case, the presence of a distinguished length scale, $L_\mathrm{exp}$, leads to the emergence of plateaus of homogenized surface potential vorticity surrounded by kinetic energy ribbons \citep{arbic_coherent_2003}. Both of these features can be seen in figure \ref{F-model_runs_mixed}.
	
	 The $k^{-3}$ surface kinetic energy spectrum \eqref{eq:equiv_KE} is only expected to hold at horizontal scales larger than $2\,\pi \, \sigma_0 \, h_\mathrm{exp}$; at smaller scales we should recover the $k^{-5/3}$ spectrum expected from uniformly stratified surface quasiogeostrophic theory.  Figure \ref{F-model_runs_mixed}(i) shows that there is indeed a steepening of the kinetic energy spectrum at horizontal scales larger than 20 km, although the model spectrum is somewhat steeper than the predicted $k^{-3}$. Similarly, although the spectrum flattens at smaller scales, the small-scale spectrum is also slightly steeper than the predicted $k^{-5/3}$.
	
	We can also examine the spectral transfer functions of $\Psc(k)$ and $\Ksc(k)$. At large scales, we expect an inertial range in surface kinetic energy, so $\Pi_K(k) = $ constant, whereas at small scales, we expect an inertial range in surface potential enstrophy, so $\Pi_P(k)= $ constant. However, figure \ref{F-model_runs_transfer} shows that although both $\Pi_K(k)$ and $\Pi_P(k)$ become approximately flat at small scales, we observe significant deviations at larger scales.
	
	\subsection{More general stratification profiles}
	
	These three idealized cases provide intuition for how the inversion function behaves for an arbitrary stratification profile, $\sigma(z)$. Generally, if $\sigma(z)$ is decreasing over some depth, then the inversion function will steepen to a super linear wavenumber dependence over a range of horizontal wavenumber whose vertical structure function significantly impinges on these depths. A larger difference in stratification between these depths leads to a steeper inversion function.
	 Analogously, if $\sigma(z)$ is increasing over some depth, then the inversion function will flatten to a sublinear wavenumber dependence, with a larger difference in stratification leading to a flatter inversion function.  
	Finally, if $\sigma(z)$ is much smaller at depth than near the surface, the inversion function will flatten to become approximately constant, and we recover an equivalent barotopic like regime, similar to the exponentially stratified example.
		
	\section{Application to the ECCOv4 ocean state estimate}\label{S-ECCO}
	
	We now show that, over the mid-latitude North Atlantic, the inversion function is seasonal at horizontal scales between 1-100 km, transitioning from $m(k) \sim k^{3/2}$ in winter to $m(k)\sim k^{1/2}$ in summer.	To compute the inversion function $m(k)$, we obtain the stratification profile $\sigma(z)=N(z)/f$ at each location from the Estimating the Circulation and Climate of the Ocean version 4 release 4 \citep[ECCOv4,][]{forget_ecco_2015} state estimate. We then compute $\Psi_k(z)$ using the vertical structure equation \eqref{eq:Psi_equation} and then use the definition of the inversion function \eqref{eq:mk} to obtain $m(k)$ at each wavenumber $k$.
		
	\subsection{The three horizontal length-scales}
	
	In addition to $L_\mathrm{mix}$ and $L_\mathrm{pyc}$ [defined in equations \eqref{eq:mixed_length} and \eqref{eq:therm_length}], we introduce the  horizontal length scale, $L_H$, the full-depth horizontal scale, defined by
	\begin{equation}\label{eq:LH}
		L_H = 2 \, \pi \, \sigma_\mathrm{ave} \, H,
	\end{equation}
	where $\sigma_\mathrm{ave}$ is the vertical average of $\sigma$ and $H$ is the local ocean depth. The bottom boundary condition becomes important to the dynamics at horizontal scales larger than $\approx L_H$.
	
	\begin{figure*}
		\centerline{\includegraphics[width=37pc]{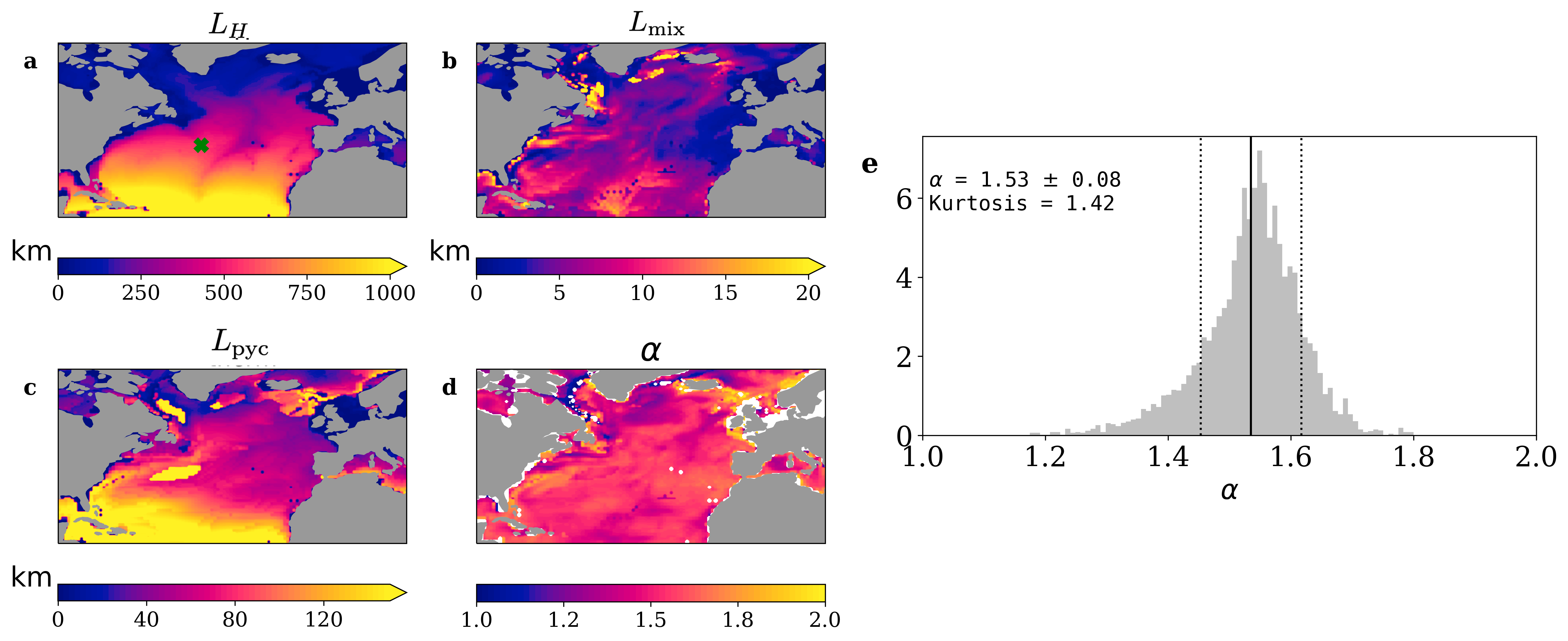}}
		\caption{Panels (a), (b), and (c) show the horizontal length scales $L_H$, $L_\textrm{mix}$, and $L_\textrm{pyc}$ as computed from 2017 January mean ECCOv4 stratification profiles, $\sigma(z) = N(z)/f$, over the North Atlantic. The green `x' in panel (a) shows the location chosen for the inversion functions of figure \ref{F-mk_realistic_small} and the model simulations of figure \ref{F-model_runs_ecco}. Panel (d) shows $\alpha$, defined by $m(k)/k^{\alpha} \approx \mathrm{constant}$, over the North Atlantic. We compute $\alpha$ by fitting a straight line to a log-log plot of $m(k)$ between $2\pi/L_\mathrm{mix}$ and $2\pi/L_\mathrm{pyc}$. Panel (e) is a histogram of the computed values of $\alpha$ over the North Atlantic. We exclude from this histogram grid cells with $L_H < 150$ km; these are primarily continental shelves and high-latitude regions. In these excluded regions, our chosen bottom boundary condition \eqref{eq:no-slip} may be influencing the computed value of $\alpha$.}
		\label{F-alpha_obs}
	\end{figure*}
	
	\begin{figure*}
		\centerline{\includegraphics[width=37pc]{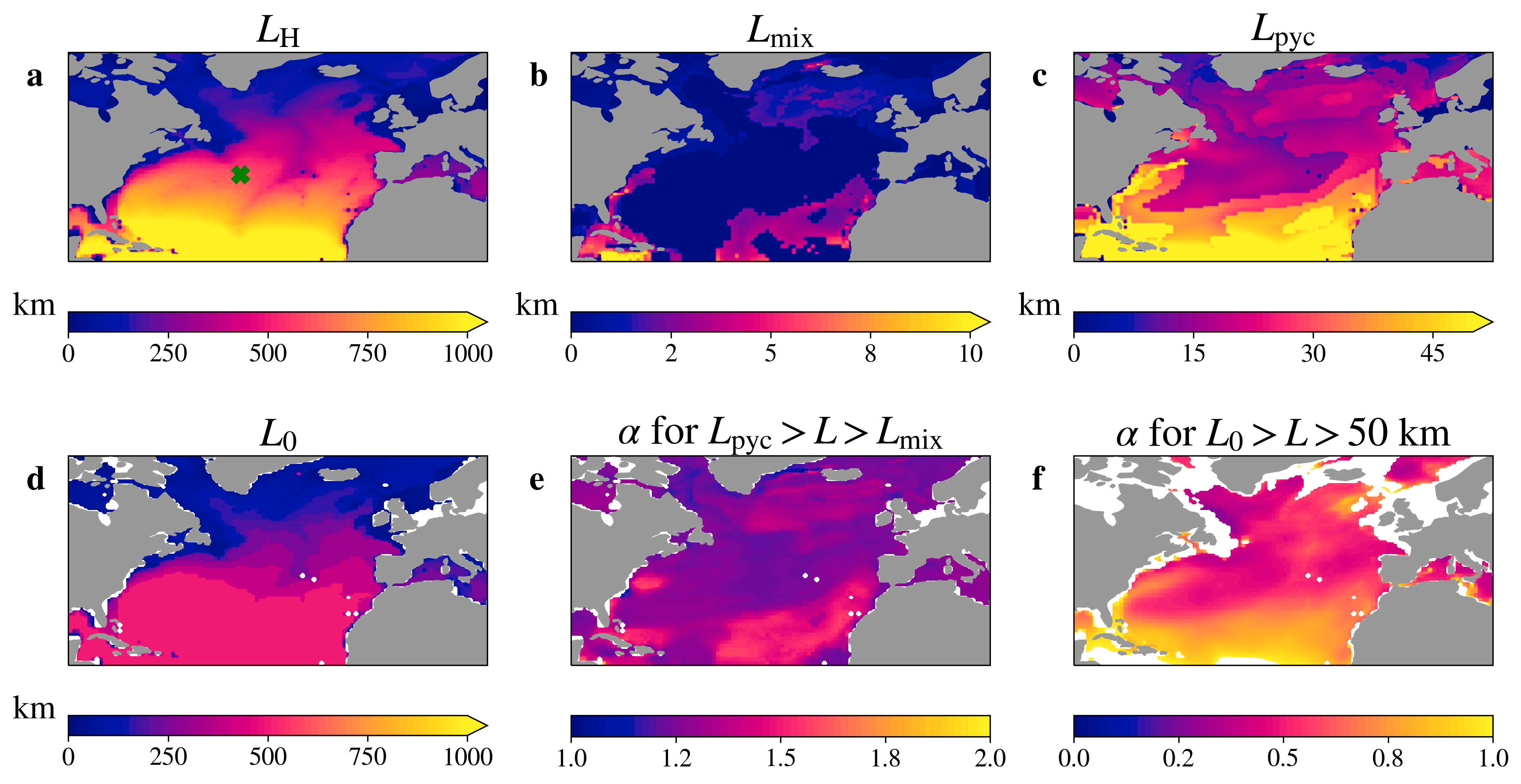}}
		\caption{Panels (a), (b), (c), and (e) are as in figure \ref{F-alpha_obs}(a)-(d), but computed from 2017 July mean stratification profiles. The calculation of $L_0$ in panel (d) is explained in the text. In panel (f), we show $\alpha$ but measured between $2\pi/(50 \, \mathrm{km})$ and $2\pi/L_0$. }
		\label{F-NorthAtlanticJul}
	\end{figure*} 
	
	We compute all three length scales using ECCOv4 stratification profiles over the North Atlantic, with results displayed in figures \ref{F-alpha_obs}(a)-(c) and \ref{F-NorthAtlanticJul}(a)-(c) for January and July, respectively. To compute the mixed-layer horizontal scale, $L_\mathrm{mix} = 2 \, \pi \sigma_0 \, h_\mathrm{mix}$, we set $\sigma_0$ equal to the stratification at the uppermost grid cell. The mixed-layer depth, $h_\mathrm{mix}$, is then defined as follows. We first define the pycnocline stratification, $\sigma_\mathrm{pyc}$, to be the maximum of $\sigma(z)$. The mixed-layer depth, $h_\mathrm{mix}$, is then the depth at which $\sigma(-h_\mathrm{mix})= \sigma_0 + \left(\sigma_\mathrm{pyc}-\sigma_0\right)/4$. Finally, the pycnocline horizontal scale, $L_\mathrm{pyc}$, is computed as $L_\mathrm{pyc}=2\, \pi \, \sigma_\mathrm{pyc} \, h_\mathrm{pyc}$, where $h_\mathrm{pyc}$ is the depth of the stratification maximum $\sigma_\mathrm{pyc}$. 

	Figures \ref{F-alpha_obs}(a) and \ref{F-NorthAtlanticJul}(a) show that $L_H$ is not seasonal, with typical mid-latitude open ocean values between $400-700$ km. On continental shelves, as well as high-latitudes, $L_H$ decreases to values smaller than $200$ km. As we approach the equator, the full-depth horizontal scale $L_H$ becomes large due to the smallness of the Coriolis parameter.
	
	Constant stratification surface quasigeostrophic theory is only valid at horizontal scales smaller than $L_\mathrm{mix}$. Figure \ref{F-alpha_obs}(b) shows that the wintertime $L_\mathrm{mix}$ is spatially variable with values ranging between $1-15$ km. In contrast, figure \ref{F-NorthAtlanticJul}(b) shows that the summertime $L_\mathrm{mix}$ is less than 2 km over most of the midlatitude North Atlantic.
	
	Finally, we expect to observe a superlinear inversion function for horizontal scales between $L_\mathrm{mix}$ and $L_\mathrm{pyc}$. The latter, $L_\mathrm{pyc}$, is shown in figures \ref{F-alpha_obs}(c) and \ref{F-NorthAtlanticJul}(c). Typical mid-latitude values range between $70-110$ km in winter but decrease to $15-30$ km in summer. In figure \ref{F-alpha_obs}(c), there is a region close to the Gulf stream with anomalously high values of $L_\mathrm{pyc}$. In this region, the stratification maximum, $\sigma_\mathrm{pyc}$ is approximately half as large as the surrounding region, but its depth, $h_\mathrm{pyc}$, is about five times deeper, resulting in larger values of $L_\mathrm{pyc}=2\,\pi \,\sigma_\mathrm{pyc}\, h_\mathrm{pyc}$.
		
	\subsection{The inversion function at a single location}
	Before computing the inversion function over the North Atlantic, we focus on a single location. However, we must first address what boundary conditions to use in solving the vertical structure equation $\eqref{eq:Psi_equation}$ for $\Psi_k(z)$. We cannot use the infinite bottom boundary condition \eqref{eq:Psi_lower} because the ocean has a finite depth. However,  given that figures \ref{F-alpha_obs}(a) and \ref{F-NorthAtlanticJul}(a) show that the bottom boundary condition should not effect the inversion function at horizontal scales smaller than 400 km in the mid-latitude North Atlantic open ocean, we choose to use the no-slip bottom boundary condition
	\begin{align}\label{eq:no-slip}
		\Psi_k(-H) = 0.
	\end{align}
	The alternate free-slip boundary condition
	\begin{align}\label{eq:free-slip}
		\d{\Psi_k(-H)}{z} = 0
	\end{align}
	gives qualitatively identical results for horizontal scales smaller than 400 km, which are the scales of interest in this study [see appendix A for the large-scale limit of $m(k)$ under these boundary conditions].\footnote{ The no-slip boundary condition \eqref{eq:no-slip} is appropriate over strong bottom friction \citep{arbic_baroclinically_2004} or steep topography \citep{lacasce_prevalence_2017} whereas the free-slip boundary condition \eqref{eq:free-slip} is appropriate over a flat bottom.}
	
	\begin{figure*}
	\centerline{\includegraphics[width=33pc]{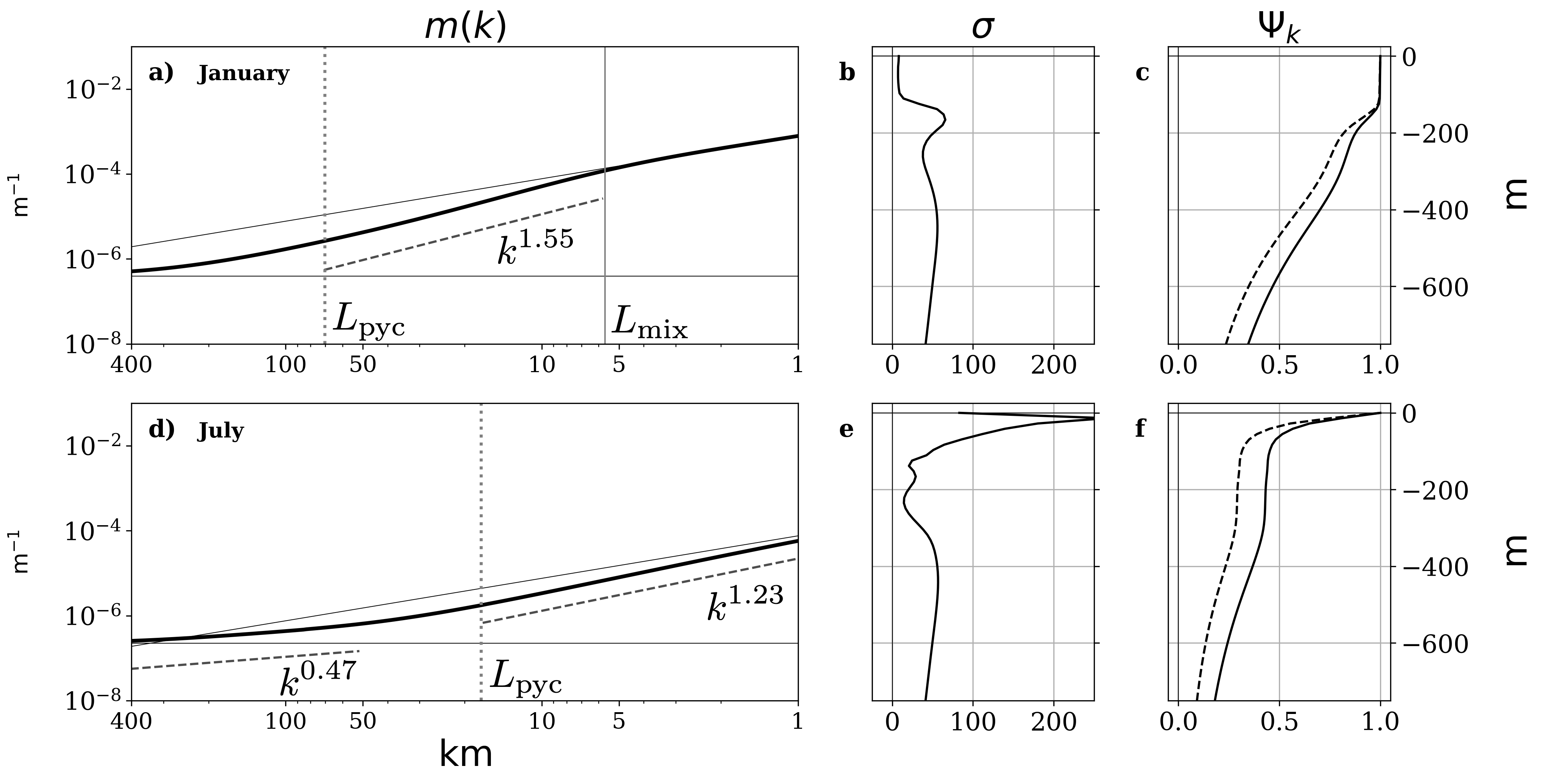}}
	\caption{As in figure \ref{F-mk_constlinconst} but for the mid-latitude North Atlantic location  ($38^\circ$ N, $45^\circ$ W)  in January [(a)-(c)] and July [(d)-(f)]. This location is marked by a green `x' in figure \ref{F-alpha_obs}(a). Only the upper 750 m of the stratification profiles and vertical structures are shown in panels (b), (c), (e) and (f).}
	\label{F-mk_realistic_small}
  	\end{figure*}

  	\begin{figure*}
	\centerline{\includegraphics[width=27pc]{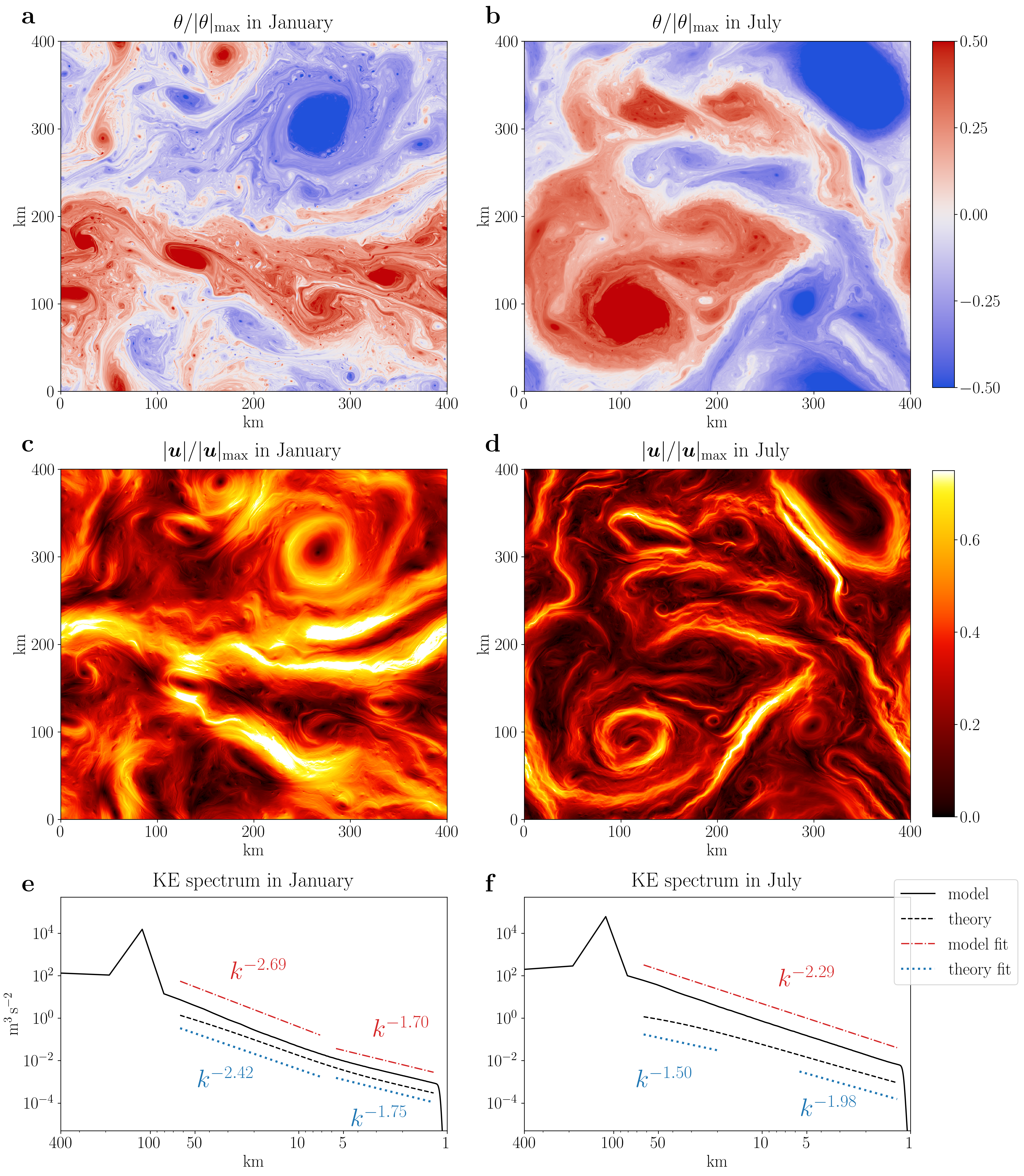}}
	\caption{Two pseudo-spectral simulations differing only in the chosen stratification profile $\sigma(z) = N(z)/f$. Both simulations use a monthly averaged 2017 stratification at the mid-latitude North Atlantic location ($38^\circ$ N,$45^\circ$ W) [see the green 'x' in figure \ref{F-alpha_obs}a] in January [(a), (c), (e)] and July [(b), (d), (f)]. The stratification profiles are obtained from the Estimating the Circulation and Climate of the Ocean version 4 release 4 \citep[ECCOv4,][]{forget_ecco_2015} state estimate. Otherwise as in figure \ref{F-model_runs_mixed}.}
	\label{F-model_runs_ecco}
 	\end{figure*}
	
\begin{figure*}
	\centerline{\includegraphics[width=33pc]{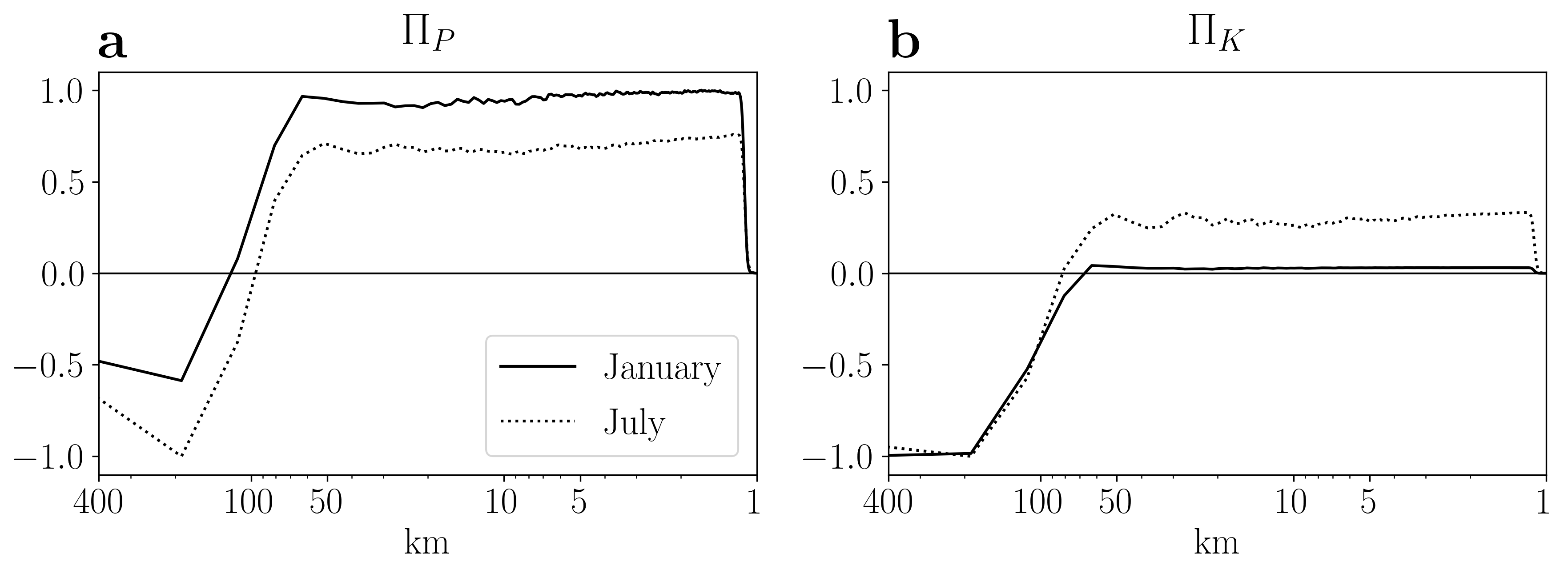}}
	\caption{The spectral density transfer functions for surface potential enstrophy (a) and surface kinetic energy (b) normalized by their absolute maximum for the two simulations in figure \ref{F-model_runs_ecco}.}
	\label{F-model_runs_real_transfer}
	\end{figure*}

	Figure \ref{F-mk_realistic_small} shows the computed inversion function in the mid-latitude North Atlantic at ($38^\circ$ N, $45^\circ$ W)  [see the green `x' in figure \ref{F-alpha_obs}(a)]. In winter, at horizontal scales smaller than $L_\mathrm{mix} \approx 5$ km, we recover the linear $m(k) \approx k/\sigma_0$ expected from constant stratification surface quasigeostrophic theory. However, for horizontal scales between  $L_\mathrm{mix} \approx 5$ km and $L_\mathrm{pyc} \approx 70$ km, the inversion function, $m(k)$, becomes as steep as a $k^{3/2}$ power law. Figure \ref{F-model_runs_ecco} shows a snapshot of the surface potential vorticity and the geostrophic velocity from a surface quasigeostrophic model using the wintertime inversion function. The surface potential vorticity snapshot is similar to the idealized mixed-layer snapshot of figure \ref{F-model_runs_mixed}(a), which is also characterized by $\alpha \approx 3/2$ (but at horizontal scales between 7-50 km). Both simulations exhibit a preponderance of small-scale vortices as well as thin filaments of surface potential vorticity. As expected, the kinetic energy spectrum [figure \ref{F-model_runs_ecco}(e)] transitions from an $\alpha\approx3/2$ regime to an $\alpha = 1$ regime near $L_\mathrm{mix} = 5$ km. Moreover, as shown in figure \ref{F-model_runs_real_transfer}, an approximate inertial range is evident between the forcing and dissipation scales.	
	
	In summer, the mixed-layer horizontal scale, $L_\mathrm{mix}$, becomes smaller than 1 km and the pycnocline horizontal scale, $L_\mathrm{pyc}$, decreases to 20 km. We therefore obtain a super linear regime, with $m(k)$ as steep as $k^{1.2}$, but only for horizontal scales between 1-20 km. Thus, although there is a range of wavenumbers for which $m(k)$ steepens to a super linear wavenumber dependence in summer, this range of wavenumbers is narrow, only found at small horizontal scales, and the steepening is much less pronounced than in winter. 
	At horizontal scales larger than $L_\mathrm{pyc}$, the summertime inversion function flattens, with the $m(k)$ increasing like a $k^{1/2}$ power law between 50-400 km. This flattening is due to the largely decaying nature of ocean stratification below the stratification maximum.

	As expected from a simulation with a sublinear inversion function at large scales, the surface potential vorticity appears spatially diffuse [figure \ref{F-model_runs_ecco}(d)] and comparable to the $\sigma_0 > \sigma_\mathrm{pyc}$ and the exponential simulations [figure \ref{F-model_runs_mixed}(b)-(c)]. However, despite having a sublinear inversion function, the July simulation is dynamically more similar to the exponential simulation rather than the  $\sigma_0 > \sigma_\mathrm{pyc}$ simulation. The July simulation displays approximately homogenized regions of surface potential vorticity surrounded by surface kinetic energy ribbons. As a result, the surface kinetic energy spectrum does not follow the predicted spectrum \eqref{eq:forward_KE}.

	\subsection{The inversion function over the North Atlantic}
	
	We now present power law approximations to the inversion function $m(k)$ over the North Atlantic in winter and summer. In winter, we obtain the power $\alpha$, where $m(k)/k^\alpha \approx \mathrm{constant}$, by fitting a straight line to $m(k)$ on a log-log plot between $2\pi/L_\mathrm{mix}$ and $2\pi/L_\mathrm{pyc}$. A value of $\alpha =1$ is expected for constant stratification surface quasigeostrophic theory. A value of $\alpha = 2$ leads to an inversion relation similar to two-dimensional barotropic dynamics. 
	 However, in general, we emphasize that $\alpha$ is simply a crude measure of how quickly $m(k)$ is increasing; we do not mean to imply that $m(k)$ in fact takes the form of a power law. Nevertheless, the power $\alpha$ is useful because, as $\alpha$-turbulence suggests (and the simulations in section \ref{S-idealized} confirm), the rate of increase of the inversion function measures the spatial locality of the resulting flow. 

	\begin{figure}
	\centerline{\includegraphics[width=19pc]{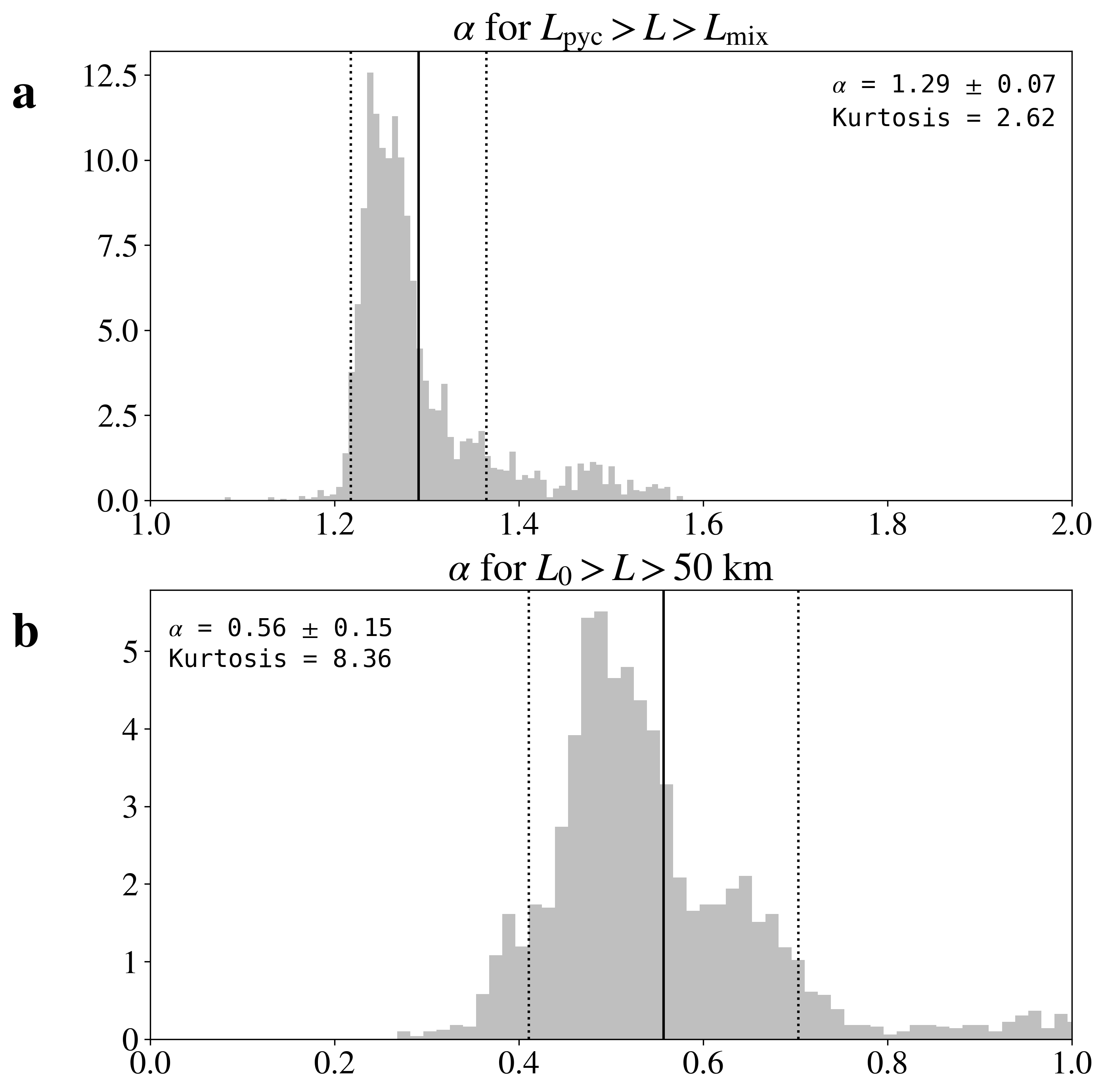}}
	\caption{Panel (a) is as in figure \ref{F-alpha_obs}(e), but with the additional restriction that $L_H < 750$ km to filter out the non-seasonal equatorial region. In panel (b), we instead plot $\alpha$ as obtained by fitting a straight line to a log-log plot of $m(k)$ between $2\pi/(50 \, \mathrm{km})$ and $2\pi/L_0$ with the same restrictions as in panel (a).}
	\label{F-alpha_dist_jul}
	\end{figure}
	
	Figure \ref{F-alpha_obs}(d) shows that we generally have $\alpha \approx 3/2$ in the wintertime open ocean. Deviations appear at high-latitudes (e.g., the Labrador sea and southeast of Greenland) and on  continental shelves where we find regions of low $\alpha$. However, both of these regions have small values of $L_H$ so that our chosen no-slip bottom boundary condition \eqref{eq:no-slip} may be influencing the computed $\alpha$ there. 
	
	A histogram of the computed values of $\alpha$ [figure \ref{F-alpha_obs}(e)] confirms that $\alpha \approx 1.53 \pm 0.08$ in the wintertime mid-latitude open ocean. This histogram only includes grid cells with $L_H >150$ km, which ensures that the no-slip bottom boundary condition \eqref{eq:no-slip} is not influencing the computed distribution.
	
	An inversion function of $m(k) \sim k^{3/2}$ implies a surface kinetic energy spectrum of $k^{-4/3}$ upscale of small-scale forcing [equation \eqref{eq:inverse_KE}] and a spectrum of $k^{-7/3}$ downscale of large-scale forcing [equation \eqref{eq:forward_KE}].  
	As we expect wintertime surface buoyancy anomalies to be forced both by large-scale baroclinic instability and by small-scale mixed-layer baroclinic instability, the realized surface kinetic energy spectrum should be between $k^{-4/3}$ and $k^{-7/3}$. Such a prediction is consistent with the finding that wintertime North Atlantic geostrophic surface velocities are mainly due to surface buoyancy anomalies \citep{lapeyre_what_2009,gonzalez-haro_global_2014}  and observational evidence of a $k^{-2}$ wintertime spectrum \citep{callies_seasonality_2015}. 
	
	 The universality of the $m(k) \sim k^{3/2}$ regime over the mid-latitudes is expected because it arises from a mechanism universally present over the mid-latitude ocean in winter; namely, the deepening of the mixed-layer. However, a comment is required on why this regime also appears at low latitudes where we do not observe deep wintertime mixed-layers. At low latitudes, the $m(k) \sim k^{3/2}$ regime emerges because there is a large scale-separation between $L_\mathrm{mix}$ and $L_\mathrm{pyc}$. The smallness of the low latitude Coriolis parameter $f$ cancels out the shallowness of the low latitude mixed-layer depth resulting in values of $L_\mathrm{mix}$ comparable to the remainder of the mid-latitude North Atlantic, as seen in figure \ref{F-alpha_obs}(b). However, no similar cancellation occurs for $L_\mathrm{pyc}$ which reaches values of $\approx 500$ km due to the smallness of the Coriolis parameter $f$ at low latitudes. As a consequence, there is a non-seasonal $m(k)\sim k^{3/2}$ regime at low latitudes for horizontal scales between $10-500$ km.
	 
	 The analogous summertime results are presented in figure \ref{F-NorthAtlanticJul}(e) and figure \ref{F-alpha_dist_jul}(a). Near the equator, we obtain values close to $\alpha \approx 3/2$, as expected from the weak seasonality there. In contrast, the midlatitudes generally display $\alpha \approx 1.2-1.3$ but this superlinear regime is only present at horizontal scales smaller than  $L_\mathrm{pyc} \approx 20-30$ km. Figure \ref{F-alpha_dist_jul}(a) shows a histogram of the measured $\alpha$ values but with the additional restriction that $L_H < 750$ km to filter out the near equatorial region (where $\alpha \approx 3/2$). 
	 	 
	The summertime inversion function shown in figure \ref{F-mk_realistic_small}(d) suggests that the inversion function flattens at horizontal scales larger than 50 km, with $m(k)$ increasing like a $k^{1/2}$ power law.
	  We now generalize this calculation to the summertime midlatitude North Atlantic by fitting a straight line to $m(k)$ on a log-log plot between $2\pi/(50 \, \mathrm{km})$ and $2\pi/L_0$ where $L_0$ is defined by
	 \begin{equation}
	 	m\left(\frac{2\,\pi}{L_0} \right) = m_0 = \left[ \int_{-H}^0  \sigma^2(s) \di s \right] ^{-1}
	 \end{equation}
	 and $m_0$ is defined by the second equality. In this case, we solve for $m(k)$ using the free-slip boundary condition \eqref{eq:free-slip}. We made this choice because $m(k)$ must cross $m_0$ in the large-scale limit if we apply the free-slip boundary condition \eqref{eq:free-slip}. In contrast, $m(k)$ asymptotes to $m_0$ from above if we apply the no-slip boundary condition \eqref{eq:no-slip}. See appendix A for more details. In any case, if we use the free-slip boundary condition \eqref{eq:free-slip}, then  $L_0$ is a horizontal length scale at which the flattening of $m(k)$ ceases and $m(k)$ instead begins to steepen in order to attain the required $H\, k^2$ dependence at large horizontal scales [see equation \eqref{eq:mk_free}]. Over the mid-latitudes North Atlantic, $L_0$ has typical values of 200-500 km [figure \ref{F-NorthAtlanticJul}(d)].
	 
	 When $\alpha$ is measured between 50 km and $L_0$, we find typical midlatitude values close to $\alpha \approx 1/2$ [figure \ref{F-NorthAtlanticJul}(f)]. A histogram of these $\alpha$ values is provided in figure \ref{F-alpha_dist_jul}(b), where we only consider grid cells satisfying $L_H > 150$ km and $L_H < 750$ km (the latter condition filters out near equatorial grid cells). The distribution is broad with a mean of $\alpha = 0.56 \pm 0.15$ and a long tail of $\alpha > 0.8$ values. Therefore, $m(k)$ flattens considerably in response to the decaying nature of summertime upper ocean stratification. It is not clear, however, whether the resulting dynamics will be similar to the $\sigma_0 > \sigma_\mathrm{pyc}$ case or the exponentially stratified case in section \ref{S-idealized}. As we have seen, the summertime simulation (in figure \ref{F-model_runs_ecco}) displayed characteristics closer to the idealized exponential case than the $\sigma_0 > \sigma_\mathrm{pyc}$ case. Nevertheless, the low summertime values of $\alpha$ indicate that buoyancy anomalies generate shorter range velocity fields in summer than in winter.
	 
	 \cite{isern-fontanet_transfer_2014} and \cite{gonzalez-haro_ocean_2020} measured the inversion function empirically, through equation \eqref{eq:transfer_func}, and found that the inversion function asymptotes to a constant at large horizontal scales (270 km near the western coast of Australia and 100 km in the Mediterranean Sea). They suggested this flattening is due to the dominance of the interior quasigeostrophic solution at large scales \citep[a consequence of equation 29 in][]{lapeyre_dynamics_2006}. 
	 We instead suggest this flattening is intrinsic to surface quasigeostrophy. In our calculation, the inversion function does not become constant at horizontal scales smaller than 400 km. However, if the appropriate bottom boundary condition is the no-slip boundary condition \eqref{eq:no-slip}, then the inversion asymptotes to a constant value at horizontal scales larger than $L_H$ (appendix A).

	 \section{Discussion and conclusion}\label{S-dicussion}
	 
	 As reviewed in the introduction, surface geostrophic velocities over the Gulf Stream, the Kuroshio, and the Southern Ocean are primarily induced by  surface buoyancy anomalies in winter \citep{lapeyre_what_2009, isern-fontanet_diagnosis_2014, gonzalez-haro_global_2014, qiu_reconstructability_2016, miracca-lage_can_2022}.
	  However, the kinetic energy spectra found in observations and numerical models are too steep to be consistent with uniformly stratified surface quasigeostrophic theory \citep{blumen_uniform_1978,callies_interpreting_2013}.
	  By generalizing surface quasigeostrophic theory to account for variable stratification, we have shown that surface buoyancy anomalies can generate a  variety of dynamical regimes depending on the stratification's vertical structure. 
	  Buoyancy anomalies generate longer range velocity fields over decreasing stratification [$\sigma'(z)\leq0$] and shorter range velocity fields over increasing stratification  [$\sigma'(z)\geq0$]. As a result, the surface kinetic energy spectrum is steeper over decreasing stratification than over increasing stratification. An exception occurs if there is a large difference between the surface stratification and the deep ocean stratification (as in the exponential stratified example of section \ref{S-idealized}). In this case, we find regions of approximately homogenized surface buoyancy surrounded by kinetic energy ribbons \citep[similar to][]{arbic_coherent_2003} and this spatial reorganization of the flow results in a steep kinetic energy spectrum. By applying the variable stratification theory to the wintertime North Atlantic and assuming that mixed-layer instability acts as a narrowband small-scale surface buoyancy forcing, we find that the theory predicts a surface kinetic energy spectrum between $k^{-4/3}$ and $k^{-7/3}$, which is consistent with the observed wintertime $k^{-2}$ spectrum \citep{sasaki_impact_2014,callies_seasonality_2015,vergara_revised_2019}. There remains the problem that mixed-layer instability may not be localized at a certain horizontal scale but is forcing the surface flow at a wide range of scales \citep{khatri_role_2021}. In this case we suggest that the main consequence of this broadband forcing is again to flatten the $k^{-7/3}$ spectrum.	  
	  
	  Over the summertime North Atlantic, buoyancy anomalies generate a more local velocity field and the surface kinetic energy spectrum is flatter than in winter. This contradicts the $k^{-3}$ spectrum found in observations and numerical models \citep{sasaki_impact_2014,callies_seasonality_2015}. However, observations also suggest that the surface geostrophic velocity is no longer dominated by the surface buoyancy induced contribution, suggesting the importance of interior potential vorticity for the summertime surface velocity \citep{gonzalez-haro_global_2014,miracca-lage_can_2022}. As such, the surface kinetic energy predictions of the present model, which neglects interior potential vorticity, are not valid over the summertime North Atlantic.
	  
	
	The situation in the North Pacific is broadly similar to that in the North Atlantic. In the Southern Ocean, however, the weak depth-averaged stratification leads to values of $L_H$ close to 150-200 km. As such, the bottom boundary becomes important at smaller horizontal scales than in the North Atlantic. Regardless of whether the appropriate bottom boundary condition is no-slip \eqref{eq:no-slip} or free-slip \eqref{eq:free-slip}, in both cases, the resulting inversion function implies a steepening to a $k^{-3}$ surface kinetic energy spectrum (appendix A). The importance of the bottom boundary in the Southern Ocean may explain the observed steepness of the surface kinetic energy spectra [between $k^{-2.5}$ to $k^{-3}$ \citep{vergara_revised_2019}] even though the surface geostrophic velocity seems to be largely due to surface buoyancy anomalies throughout the year \citep{gonzalez-haro_global_2014}.
	
	The claims made in this article can be explicitly tested in a realistic high-resolution ocean model; this can be done by finding regions where the surface streamfunction as reconstructed from sea surface height is highly correlated to the surface streamfunction as reconstructed from sea surface buoyancy \citep[or temperature, as in][]{gonzalez-haro_global_2014}. Then, in regions where both streamfunctions are highly correlated, the theory predicts that the inversion function, as computed from the stratification [equation \eqref{eq:mk}], should be identical to the inversion function computed through the surface streamfunction and buoyancy fields [equations \eqref{eq:transfer_func} and \eqref{eq:transfer_inversion}]. Moreover, in these regions, the model surface kinetic energy spectrum must be between the inverse cascade and forward cascade kinetic energy spectra [equations \eqref{eq:inverse_KE} and \eqref{eq:forward_KE}].
		
	 Finally the vertical structure equation \eqref{eq:Psi_equation} along with the inversion relation \eqref{eq:theta_fourier} between $\thetak$ and $\psik$ suggest the possibility of measuring the buoyancy frequency's vertical structure, $N(z)$, using satellites observations. This approach, however, is limited to regions where the surface geostrophic velocity is largely due to surface buoyancy anomalies. By combining satellite measurements of sea surface temperature and sea surface height, we can use the inversion relation \eqref{eq:theta_fourier} to solve for the inversion function. Then we obtain $N(z)$ by solving the inverse problem for the vertical structure equation \eqref{eq:Psi_equation}. How practical this approach is to measuring the buoyancy frequency's vertical structure remains to be seen.

\acknowledgments

We sincerely thank Bob Hallberg, Isaac Held, and Sonya Legg for useful comments on earlier drafts of this manuscript. We also thank Guillaume Lapeyre and an anonymous reviewer whose comments and recommendations greatly helped with our presentation. 
This report was prepared by Houssam Yassin under award NA18OAR4320123 from the National Oceanic and Atmospheric Administration, U.S. Department of Commerce. The statements, findings, conclusions, and recommendations are those of the authors and do not necessarily reflect the views of the National Oceanic and Atmospheric Administration, or the U.S. Department of Commerce. 

\datastatement

The ECCO data \citep{ecco_dataset_2021,ecco_synopsis_2021} is available from NASA’s Physical Oceanography Distributed Active Archive Center (https://podaac.jpl.nasa.gov/ECCO). The \texttt{pyqg} model is available on Github (https://github.com/pyqg/pyqg).

\appendix[A]
\appendixtitle{The small- and large-scale limits}\label{A-small_large}
	   	 
	\subsection{The small-scale limit}
	Let $h$ be a characteristic vertical length scale associated with $\sigma(z)$ near $z=0$. Then, in the small-scale limit, $k\, \sigma_0 \, h \gg 1$, the infinite bottom boundary condition \eqref{eq:Psi_lower} is appropriate.
	With the substitution
	\begin{align}
		\Psi(z) = \sigma(z) \, P(z),
	\end{align}
	we transform the vertical structure equation \eqref{eq:Psi_equation} into a Schrödinger equation,
	\begin{equation}\label{eq:app-P_equation}
		\dd{P}{z} = \left[- \frac{1}{\sigma}\dd{\sigma}{z} + 2 \left(\frac{1}{\sigma}\d{\sigma}{z}\right)^2 + k^2 \, \sigma^2 \right] P,
	\end{equation}
	with a lower boundary condition
	\begin{equation}
		\sigma \, P \rightarrow 0 \quad \text{ as } \quad z\rightarrow -\infty.
	\end{equation} 
	In the limit $k\, \sigma_0 \, h \gg 1$, the solution to the Schrödinger equation equation \eqref{eq:app-P_equation} is given by
	 \begin{equation}
	  	\Psi_k(z) \approx \sqrt{\frac{\sigma(z)}{\sigma_0}} \, \exp\left({k\, \int_0^z \sigma(s) \di{s}}\right).
	  \end{equation}
	  On substituting $\Psi_k(z)$ into the definition of the inversion function \eqref{eq:mk}, we obtain $m(k) \approx k/\sigma_0$ to leading order in $(k\sigma_0 h)^{-1}$. Therefore, the inversion relation in the small-scale limit coincides with the familiar inversion relation of constant stratification surface quasigeostrophic theory \citep{blumen_uniform_1978,held_surface_1995}.
	  
	 \subsection{The large-scale free-slip limit}
	Let $k_H= 2\pi/L_H$, where the horizontal length scale $L_H$ is defined in equation \eqref{eq:LH}. Then, in the large-scale limit, $k/k_H\ll 1$, we assume a solution of the form
	\begin{equation}\label{eq:Psi_series_large}
		\Psi_k(z) = \Psi_k^{(0)}(z) + \left(\frac{k}{ k_H}\right)^2 \, \Psi_k^{(1)}(z) + \cdots.
	\end{equation}
	Substituting the series expansion \eqref{eq:Psi_series_large} into the vertical structure equation \eqref{eq:Psi_equation} and applying the free-slip bottom boundary condition \eqref{eq:free-slip} yields
	\begin{equation}\label{eq:Psi_free_sol}
		\Psi_k(z) \approx A \left[1 + k^2 \int_{-H}^z \sigma^2(s) \, \left(s + H\right) \di\,s + \cdots \right],
	\end{equation}
	where $A$ is a constant determined by the upper boundary condition \eqref{eq:Psi_upper}.
	To leading order in $k/k_H$, the large-scale vertical structure is independent of depth.
	
	Substituting the solution \eqref{eq:Psi_free_sol} into the definition of the inversion function \eqref{eq:mk} gives
	\begin{equation}\label{eq:mk_free}
		m(k) \approx H \, k^2.
	\end{equation}
	Therefore, over a free-slip bottom boundary, the large-scale dynamics resemble two-dimensional vorticity dynamics, generalizing the result of \cite{tulloch_theory_2006} to arbitrary stratification $\sigma(z)$.

	 \subsection{The large-scale no-slip limit}

	 Substituting the expansion \eqref{eq:Psi_series_large} into the vertical structure equation \eqref{eq:Psi_equation} and applying the no-slip lower boundary condition \eqref{eq:no-slip} yields
	\begin{align}\label{eq:Psi_no_sol}
	\begin{split}
		\Psi_k &(z) \approx B \Bigg[ \int_{-H}^z \, \sigma^2(s) \, \di \, s \, +\\ &k^2 \, \int_{-H}^z \sigma^2(s_3) \int_{-H}^{s_3} \int_{-H}^{s_2} \sigma^2(s_1) \, \di s_1 \, \di s_2 \, \di s_3 \Bigg],
		\end{split}
	\end{align}
	where $B$ is a constant determined by the upper boundary condition \eqref{eq:Psi_upper}. Substituting the solution \eqref{eq:Psi_no_sol} into the definition of the inversion function \eqref{eq:mk} gives
	\begin{equation}\label{eq:mk_no}
		m(k) \approx m_1 \left(k_\sigma^2 +  k^2 \right),
	\end{equation}
	where $k_\sigma = \sqrt{m_0/m_1}$ is analogous to the deformation wavenumber, the constant $m_0$ is given by
	\begin{equation}
		m_0 = \left[ \int_{-H}^0  \sigma^2(s) \di s \right] ^{-1},
	\end{equation}
	and $m_1$ is some constant determined by integrals of $\sigma(z)$. If $\sigma(z)$ is positive then both $m_0$ and $m_1$ are also positive. Therefore, over a no-slip bottom boundary, the large-scale dynamics resemble those of the equivalent barotropic model.

\appendix[B]

\appendixtitle{Inversion function for piecewise constant stratification}

We seek a solution to the vertical structure equation \eqref{eq:Psi_equation} for the piecewise constant stratification \eqref{eq:step_strat} with upper boundary condition \eqref{eq:Psi_upper} and the infinite lower boundary condition \eqref{eq:Psi_lower}. The solution has the form 
\begin{equation}\label{eq:Psi_sharp}
	\Psi_k(z) = \cosh\left(\sigma_0 \, k\,z\right) + a_2 \sinh\left(\sigma_0 \, k\, z\right)
\end{equation}
for $-h \leq z \leq 0$, and
\begin{equation}
	\Psi_k(z) = a_3\,  e^{\sigma_\mathrm{pyc} k(z+h)}
\end{equation}
for $-\infty < z < -h$. 
To determine $a_2$ and $a_3$, we require $\Psi_k(z)$ to be continuous across $z=-h$ and that its derivative satisfy
\begin{equation}
	\frac{1}{\sigma_0^2} \, \d{\Psi_k(-h^+)}{z} = \frac{1}{\sigma_\mathrm{pyc}^2} \, \d{\Psi_k(-h^-)}{z},
\end{equation}
where the $-$ and $+$ superscripts indicate limits from the below and above respectively. Solving for $a_2$ and substituting the vertical structure \eqref{eq:Psi_sharp} into the definition of the inversion function \eqref{eq:mk} then yields $m(k)$.

\appendix[C]

\appendixtitle{The numerical model}

We solve the time-evolution equation \eqref{eq:theta_equation} using the pseudo-spectral \texttt{pyqg} model \citep{abernathey_pyqgpyqg_2019}. To take stratification into account, we use the inversion relation \eqref{eq:theta_inversion}. Given a stratification profile $\sigma(z)$ from ECCOv4, we first interpolate the ECCOv4 stratification profile with a cubic spline onto a vertical grid with 350 vertical grid points. We then numerically solve the vertical structure equation \eqref{eq:Psi_equation}, along with boundary conditions \eqref{eq:Psi_upper} and either \eqref{eq:no-slip} or \eqref{eq:free-slip}, and obtain the vertical structure at each wavevector $\vec k$. Using the definition of the inversion function \eqref{eq:mk} then gives $m(k)$.

 We apply a large-scale forcing, $F$, between the (non-dimensional) wavenumbers $3.5<k<4.5$ in all our simulations, corresponding to horizontal length scales 88 - 114 km. Otherwise, the forcing $F$ is as in \cite{smith_turbulent_2002}. The dissipation term can be written as
 \begin{equation}
     D = r_d \, \theta + \mathrm{ssd}
 \end{equation}
 where $r_d$ is a damping rate and $\mathrm{ssd}$ is small-scale dissipation. Small-scale dissipation is through an exponential surface potential enstrophy filter as in \cite{arbic_coherent_2003}. Simulations have a horizontal resolution of $1024^2$ horizontal grid points.


\bibliographystyle{ametsocV6}
\bibliography{references}

\end{document}